\documentclass[epjST,final]{svjour}

\usepackage{graphics}
\usepackage{amsmath,bm}
\usepackage{mathtools}
\usepackage{braket}
\usepackage{bbold}
\usepackage{nicefrac}
\usepackage{upgreek}

\newcommand*\Bell{\ensuremath{\boldsymbol\ell}}
\newcommand{\AR}{{\textrm{A}}}

\newcommand{\BR}{{\textrm{B}}}

\renewcommand{\makeheadbox}{\relax}
\renewcommand{\ProcessRunnHead}{\relax}

\begin{document}
\title{Long-range Rydberg molecules, Rydberg macrodimers and Rydberg aggregates in an ultracold Cs gas}
\subtitle{Investigation of long-range interactions between atoms in electronically highly excited states}
\author{Heiner Sa{\ss}mannshausen\thanks{\email{heiner.sassmannshausen@phys.chem.ethz.ch}} \and Johannes Deiglmayr \and Fr{\'e}d{\'e}ric Merkt}
\institute{Laboratory of Physical Chemistry, ETH Z{\"u}rich, Vladimir-Prelog-Weg 2, 8093 Z{\"u}rich, Switzerland  \vspace{0.5cm} \\ (Dated: April 26, 2016)}
\abstract{
We present an overview of our recent investigations of long-range interactions in an ultracold Cs Rydberg gas. These interactions are studied by high-resolution photoassociation spectroscopy, using excitation close to one-photon transitions into $n$p$_{3/2}$ Rydberg states with pulsed and continuous-wave ultraviolet laser radiation, and lead to the formation of long-range Cs$_2$ molecules. We observe two types of molecular resonances. The first type originates from the correlated excitation of two atoms into Rydberg-atom-pair states interacting at long range via multipole-multipole interactions. The second type results from the interaction of one atom excited to a Rydberg state with one atom in the electronic ground state. Which type of resonances is observed in the experiments depends on the laser intensity and frequency and on the pulse sequences used to prepare the Rydberg states. We obtain insights into both types of molecular resonances by modelling the interaction potentials, using a multipole expansion of the long-range interaction for the first type of resonances and a Fermi-contact pseudo-potential for the second type of resonances. We analyse the relation of these long-range molecular resonances to molecular Rydberg states and ion-pair states, and discuss their decay channels into atomic and molecular ions. In experiments carried out with a two-colour two-photon excitation scheme, we observe a large enhancement of Rydberg-excitation probability, which we interpret as a saturable autocatalytic antiblockade phenomenon.
} 
\maketitle
\section{Introduction}
\label{intro}
This article presents the results of high-resolution spectroscopic experiments in ultracold samples of Cs vapour. The transitions we investigate in these samples lead to the excitation of Cs atoms to Rydberg states\,\cite{gallagher_rydberg_2005} in which one electron, the Rydberg electron, orbits at large distance around the positively charged full-shell Cs$^+$ ion core. In these experiments, we are not primarily interested in the properties of isolated Cs Rydberg atoms (although we can measure them with high accuracy\,\cite{sassmannshausen2013,deiglmayr2016a}), but in the long-range interactions of these atoms with ground-state atoms and other Rydberg atoms located in their vicinity. We are particularly interested in characterising the weakly-bound molecular systems that result from these interactions.

Interactions involving high Rydberg states have been at the focus of scientific investigations for almost a century\,\cite{Amaldi1934,Fermi1934}, and improved experimental tools and techniques have continuously provided new insights, \emph{e.g.} on the mechanisms responsible for pressure shifts in Rydberg spectra\,\cite{Amaldi1934,Fermi1934,Greene2000,bendkowsky2009,bendkowsky2010,li2011,Anderson2014,Anderson2014b,sassmannshausen2015b}, on elastic and inelastic collisions in Rydberg-atom samples\,\cite{vitrant1982,gallagher1982,gallagher_rydberg_2005}, on the evolution of a dense Rydberg gas into a cold plasma\,\cite{vitrant1982,robinson2000,li2005}, on two- and many-body entanglement for quantum-information processing\,\cite{lukin2001,urban2009,Gaetan2009,wilk2010}, on the generation of unusual quantum phases\,\cite{lee2011,carr2013,malossi2014}, and on the aggregation of highly-excited atoms and molecules\,\cite{farooqi2003,stanojevic2006,stanojevic2008,schwettmann2007,overstreet2007,overstreet2009,deiglmayr2014,sassmannshausen2015}.

We focus our attention on two classes of weakly interacting molecular systems involving Rydberg states: i) long-range Rydberg molecules consisting of a Rydberg atom and a ground-state atom located within the Rydberg-electron orbit. The binding results from the (short-range) scattering interaction betweeen the Rydberg electron and the ground-state atom. The existence of metastable bound states in such systems has been predicted by Greene \textit{et al.}\,\cite{Greene2000} and confirmed experimentally by Bendkowsky \textit{et al.}\,\cite{bendkowsky2009}; ii) Rydberg-atom-pair states interacting through their electric multipole moments, which are also referred to as Rydberg macrodimers and were first observed by Farooqi \textit{et al.}\,\cite{farooqi2003}.

Weakly bound molecular systems are typically made of two or more distant constituents that are predominantly held together by long-range forces. Their physical properties depend on the nature of the constituents, \emph{e.g.} whether these are atomic or molecular species, or a nucleus, or an electron, and whether the constituents are in their electronic ground state or in an electronically excited state. They also depend on the nature of the interactions that hold the constituents together, \emph{e.g.} magnetic or electric interactions involving charges, dipoles and quadrupoles. These interactions, which are sufficiently weak not to significantly alter the properties of the constituents, give rise to a long-range potential $V_{\rm inter}(\vec{R}_{ij,i < j}; i;j=1,...,N)$. The relative motion of the $N$ constituents can be described by the Hamiltonian
\begin{equation}
 \label{eq:intro}
\hat{H} = \hat{H}_1^0 +\hat{H}_2^0 + \cdot\cdot\cdot + \hat{H}_N^0 + V_{\rm inter}(\vec{R}_{ij}),
\end{equation}
where $\hat{H}_i^0$ represents the Hamiltonian operator of the isolated $i$-th constituent and $\vec{R}_{ij}$ the position vector of constituent $j$ relative to constituent $i$.

A third factor of importance in describing the behaviour of weakly-bound systems is the amplitude of the wavefunction in short-range regions, where the potential deviates from the analytic form adequate to describe the long-range region. In pure long-range states\,\cite{stwalley1978}, this amplitude vanishes, usually because a high potential barrier separates the short- and the long-range regions. Even small amplitudes in short-range regions significantly modify the structure and dynamics of long-range systems. In molecular Rydberg states, the amplitude of the Rydberg-electron wavefunction near the ion core determines the quantum defect and the rate of decay by predissociation, autoionisation and radiative processes. In the long-range Rydberg molecules introduced above, the short-range interaction of the Rydberg electron with the ground-state atom located within the Rydberg-electron orbit determines the binding energy\,\cite{Greene2000} and the dipole moment\,\cite{Greene2000,li2011,booth2015}. In Rydberg-atom pairs, almost immediate ionisation takes place when the two Rydberg atoms approach each other by less than the LeRoy radius\,\cite{leroy1973}.
\begin{figure}[t!]
\centering
\resizebox{0.93\linewidth}{!}{%
 \includegraphics{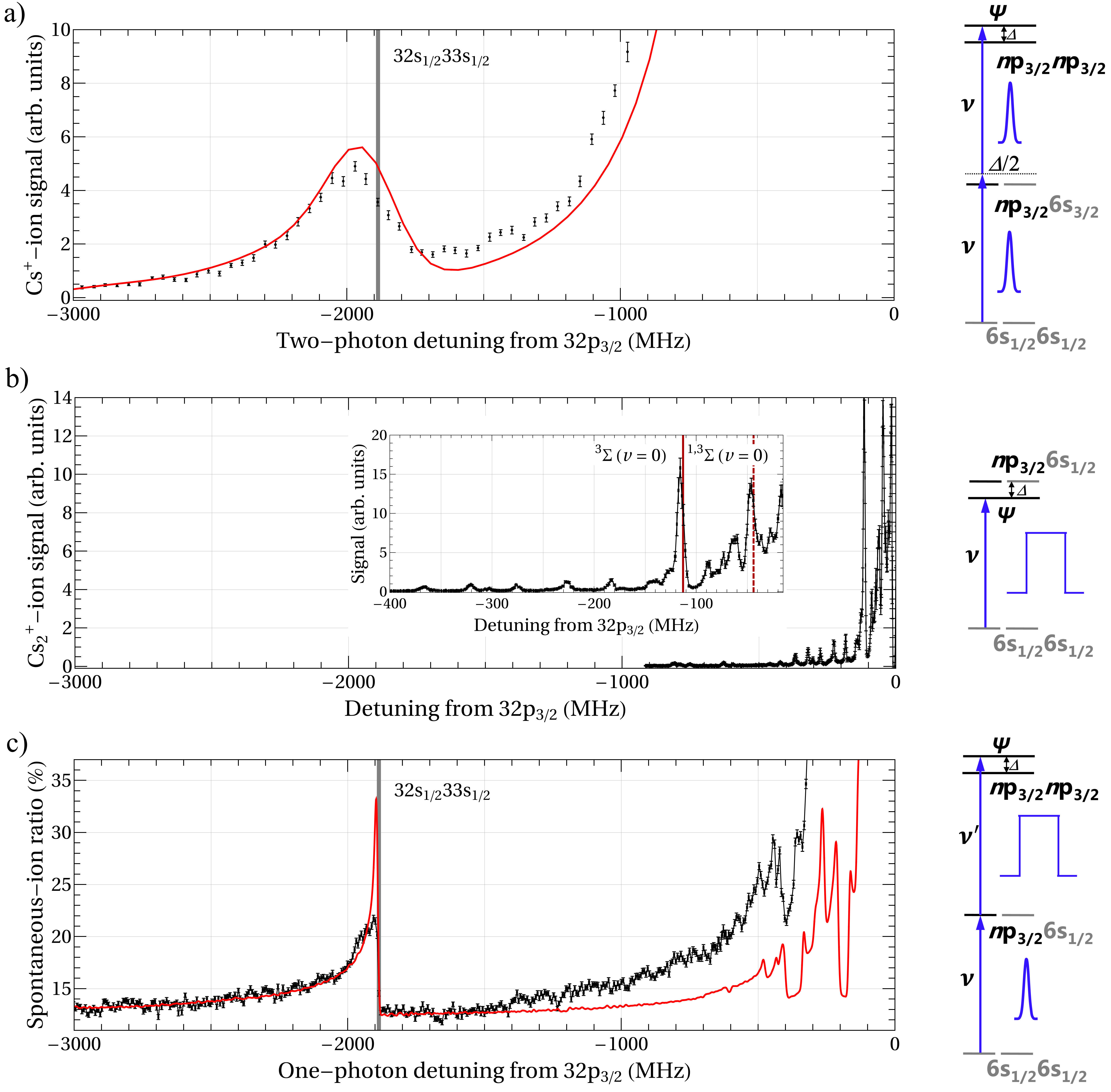}}
\caption{a) - c) Comparison of spectra measured in the same frequency range and similar samples of ultracold Cs atoms, but with different laser-excitation pulse shapes and intensities. The spectrum in a) was recorded with a 4.4-ns-long pulsed UV laser (pulse energy of $\sim10$\,$\upmu$J, peak power 2\,kW), and the vertical grey bar denotes the position of the 32s$_{1/2}33$s$_{1/2}$ dissociation asymptote. The spontaneous Cs$^+$-ion yield is plotted against the two-photon ($2 \times \nu_{\rm UV}$) detuning from the 32p$_{3/2}$32p$_{3/2}$ asymptote. The spectrum in b) was recorded with a 40\,$\upmu$s long pulse cut out of the output of a 100\,mW cw single-mode UV laser (pulse energy $\sim4$\,$\upmu$J), and the Cs$_2^+$-ion yield is shown as a function of the detuning from the 32p$_{3/2}$ Rydberg state. The inset depicts the same spectrum in the frequency range of $-400$\,MHz to $-20$\,MHz relative to the atomic resonance. The calculated positions of the $^3{\rm \Sigma} (v=0)$ and $^{1,3}{\rm \Sigma} (v=0)$ states (see section\,2 for details) are indicated by solid and dashed red lines, respectively. The spectrum in c) was recorded with a cw single-mode UV laser, after resonant excitation of a few 32p$_{3/2}$ ``seed'' Rydberg atoms. The signal corresponds to the average number of Rydberg-atom pairs per ``seed'' atom and is obtained as the ratio of Cs$^+$-ion signals resulting from the spontaneous ionisation of the sample to Cs$^+$ ions arising from the pulsed field ionisation of the seed Rydberg atoms. Red curves are simulated spectra using the model described in section\,\ref{sec:pulsed}.}
\label{fig:1}
\end{figure}

We study these weakly bound molecular states by high-resolution spectroscopy of dense samples of ultracold Cs atoms released from a magneto-optical trap (MOT). The molecular resonances are observed as weak satellite lines in Rydberg-excitation spectra in the vicinity of atomic $n$p$_{3/2} \leftarrow 6$s$_{1/2}$ transitions. Depending on the intensities, pulse lengths, frequency detunings, and bandwidths of the laser pulses, typically only one of the two afore-mentioned classes of weakly-interacting long-range molecular states is accessed. An overview of the different experimental spectra obtained under different experimental conditions is presented in Fig.\,\ref{fig:1}. When we excite the sample with an intense pulsed laser we observe two-photon transitions to Rydberg-atom-pair states (see Fig.\,\ref{fig:1}a)). The spectral widths of the resonances reflect the Fourier-transform-limited bandwidth of the short pulses which is $\sim130$\,MHz. In this case, the two-photon 32s$_{1/2}33$s$_{1/2} \leftarrow 6$s$_{1/2}6$s$_{1/2}$ transition is observed at a two-photon detuning of $2 \times 948$\,MHz from the 32p$_{3/2}32$p$_{3/2}$ dissociation asymptote. The 32s$_{1/2}33$s$_{1/2} \leftarrow 6$s$_{1/2}6$s$_{1/2}$ transition is weakly allowed because of the long-range configuration interaction of the 32s$_{1/2}33$s$_{1/2}$ pair state with the 32p$_{3/2}32$p$_{3/2}$ pair state which is optically accessible by two-photon electric-dipole transitions from the 6s$_{1/2}6$s$_{1/2}$ ground state\footnote{We use here and in the following the two-atom label ${nl}_J{n'l'}_{J'}$ (with the quantum numbers $l$ and $J$ of the orbital angular momentum and the total electronic angular momentum, respectively) to designate dissociation asymptotes and the Rydberg-atom-pair states.}. The rate of this two-photon transition is enhanced by the presence of the 32p$_{3/2}6$s$_{1/2}$ intermediate level. However, for most Rydberg-atom-pair states, the detunings from the $n$p$_{3/2}$6s$_{1/2}$ intermediate level are large and the excitation of the pair states requires high laser intensities.

The spectrum presented in Fig.\,\ref{fig:1}b) was recorded with a 40-$\upmu$s-long pulse with a bandwidth of $\sim2$\,MHz cut out of the output of a single-mode cw UV laser and also shows the spectral region around the 32p$_{3/2} \leftarrow 6$s$_{1/2}$ transition. Series of sharp molecular resonances at detunings of less than 400\,MHz from the atomic Rydberg transition are observed. These resonances are attributed to the excitation of long-range Rydberg molecules consisting of a Rydberg atom interacting with one ground-state atom by low-energy electron--neutral-atom scattering. The resonance corresponding to the two-photon excitation of the 32s$_{1/2}$33s$_{1/2}$ Rydberg-atom-pair state is not observed in this spectrum because the intensity of the UV laser is too low. The resonances resulting from the long-range Rydberg molecules are not observed with the intense UV-laser pulses used to record the spectrum shown in Fig.\,\ref{fig:1}a because their bandwidth is too large to resolve the molecular resonances and the signals are masked by larger signals that originate from the excitation of Rydberg-atom-pair states.

The 32s$_{1/2}$33s$_{1/2}$ resonance can also be studied using a low-intensity UV laser, but in resonant sequential 32s$_{1/2}33$s$_{1/2} \xleftarrow{\nu'} 6$s$_{1/2}32$p$_{3/2} \xleftarrow{\nu} 6$s$_{1/2}6$s$_{1/2}$ two-photon transitions (see Fig.\,\ref{fig:1}c). In these experiments, a few atoms are resonantly excited to 32p$_{3/2}$ Rydberg states before excitation with the narrow-band UV laser in a second resonant photoassociation step. The resulting spectrum is presented in Fig.\,\ref{fig:1}c). The spectral resolution is $\sim5$\,MHz and the pronounced asymmetry of the molecular resonance correlated to the 32s$_{1/2}33$s$_{1/2}$ dissociation asymptote is clearly revealed. The red-degraded line shape originates from the attractive nature of the molecular states.
\begin{figure}[t!]
\centering
\resizebox{0.8\linewidth}{!}{%
 \includegraphics{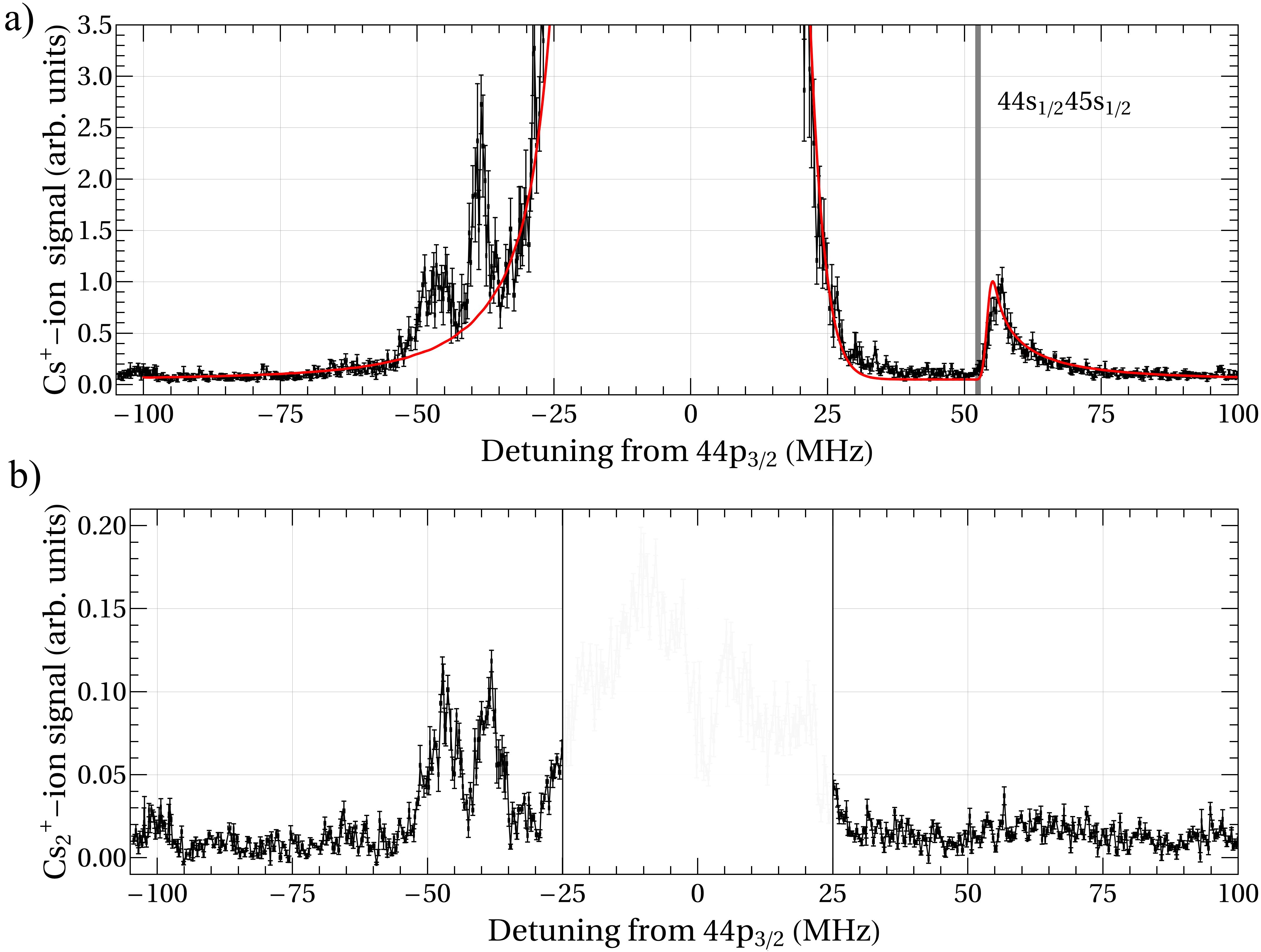}}
\caption{Spectra recorded in the vicinity of the 44p$_{3/2} \leftarrow 6$s$_{1/2}$ transition measured with the single-mode cw UV laser.  a) The spectrum shows the Cs$^+$-ion signal resulting from pulsed field ionisation. The vertical grey bar indicates the position of the 44s$_{1/2}45$s$_{1/2}$ dissociation asymptote. The red curve represents a simulated spectrum using the model described in section\,\ref{sec:pulsed}. b) The spectrum shows the signal of spontaneously produced molecular Cs$_2^+$ ions. No reliable signal could be monitored near the atomic resonance (white area between the two vertical black lines).}
\label{fig:2}
\end{figure}

At higher $n$ values, the detunings of the $n$s$_{1/2}(n+1)$s$_{1/2}$ pair states relative to the $n$p$_{3/2}n$p$_{3/2}$ states decrease and the strength of the van der Waals interaction increases significantly. At $n=44$, the asymptotes are separated by only $+105$\,MHz and the 44s$_{1/2}45$s$_{1/2} \leftarrow 6$s$_{1/2}6$s$_{1/2}$ single-colour two-photon excitation can be observed with a low-intensity single-mode UV laser. The corresponding photoassociation spectra in the vicinity of the atomic 44p$_{3/2} \leftarrow 6$s$_{1/2}$ transition are depicted in Fig.\,\ref{fig:2}. Additionally to the resonance at a detuning of $2 \times 53$\,MHz, resonances on the low-frequency side of the atomic transition are observed at detunings of $\sim-47$\,MHz and $\sim-38$\,MHz. At these detunings, the spontaneous formation of molecular Cs$_2^+$ ions is observed (see Fig.\,\ref{fig:2}b). This observation allows the assignment of these resonances to the photoassociation to long-range Rydberg molecules bound by low-energy electron--neutral-atom scattering (see Fig.\,\ref{fig:1}b and section\,2). The spectrum in Fig.\,\ref{fig:2}a is thus a unique example of the occurrence of both types of long-range molecular states in a single experiment.

The interpretation of the observed photoassociation resonances requires a detailed modelling of the relevant interactions. The red traces in Fig.\,\ref{fig:1} represent simulated spectra for the correlated excitation of interacting Rydberg-atom-pair states (see section\,3.1). In Fig.\,\ref{fig:1}a) and c), all experimentally observed resonances are faithfully reproduced by the simulation, which enables their unambiguous assignment as Rydberg-atom-pair states. The comparison between simulated and experimental spectra in Fig.\,\ref{fig:2}a confirms the different origins of the observed resonances discussed above. Indeed, the simulation only takes into account Rydberg-atom-pair resonances.

This review is organised as follows: In the next section, we discuss the long-range Rydberg molecules bound by low-energy electron--neutral-atom scattering. The third section is dedicated to studies of Rydberg-atom-pair states. The results of one- and two-colour excitation schemes are presented and compared to spectra simulated using calculated Rydberg-atom-pair potential-energy functions. We detect all molecular resonances by their ionic decay products. In section\,\ref{sec:decay}, possible decay channels leading to spontaneous ionisation of both classes of long-range molecular states are presented and discussed in relation to highly vibrationally excited Cs$_2^*$ and Cs$_2^+$ molecules and Cs$^+$Cs$^-$ ion-pair states. The article ends with a brief summary.

Whereas most of the results presented in this review article have been published, the results presented in subsection\,3.3 on resonant two-photon two-colour antiblockade effects are shown here for the first time.

\section{Long-range Rydberg molecules}
\label{sec:trilos}
Long-range molecular states supported by a Fermi-contact-type potential\,\cite{Fermi1934} describing the interaction of a Rydberg electron and a ground-state atom \textit{via} electron--neutral-atom scattering were first predicted by Greene \textit{et al.}\,\cite{Greene2000}. In these states, an atom in its electronic ground state is located inside the large electron orbit of a Rydberg atom with which it interacts \textit{via} the slow Rydberg electron. This binding mechanism does not fit into the usual categories of chemical bonds, namely covalent, ionic, metallic or van der Waals\,\cite{hogan2009}. Two main sub-classes of these long-range molecular states were already predicted in Ref.\,\cite{Greene2000}: Molecular states asymptotically correlated to low-$l$ Rydberg states with bond strengths corresponding to typically a few MHz, and states correlated to high-$l$ atomic Rydberg states with bond strengths of several GHz for Rydberg states around $n=30$. Because the binding energies are much lower than the thermal energy at room temperature, the observation of such metastable molecules is only possible through photoassociation of ultracold atoms in the gas phase. Even though the effect of the perturbation of Rydberg series by ground-state atoms was observed early through line broadenings and line shifts\,\cite{Amaldi1934}, the preparation of the long-range Rydberg molecules in well-defined quantum states requires ultracold samples. Experimentally, the long-range Rydberg molecules were first observed through photoassociation of ultracold Rb atoms using $n{\rm s}_{1/2} \leftarrow 6{\rm s}_{1/2}$ two-photon transitions\,\cite{bendkowsky2009}. Since then, they have also been observed using $n$d\,\cite{Krupp2014,Anderson2014} and $n$p\,\cite{manthey2015} Rydberg states of Rb and in Cs\,\cite{Tallant2012,sassmannshausen2015b,booth2015} and Sr\,\cite{desalvo2015}. Molecular resonances originating from electron--neutral-atom scattering were also observed in transitions from the electronic ground state of Rb$_2$ to $n$p Rydberg states\,\cite{bellos2013}. Further experiments revealed additional properties of these long-range molecules. Long-range states correlated asymptotically with $n$s$_{1/2}$ Rydberg states in Rb were found to be bound by internal quantum reflection at a steep drop of the molecular potential towards smaller internuclear separation\,\cite{bendkowsky2010}. Recently, high-$l$-mixed molecules were observed with very large induced dipole moments\,\cite{booth2015}. The lifetimes of Rb$_2$\cite{butscher2011} and Sr$_2$\cite{camargo2016} molecules have been investigated, and values similar to the parent atomic Rydberg state were found.

Most of the experiments have focused so far on the alkali-metal atoms Cs and Rb, for which singlet and triplet scattering channels for electron-atom scattering exist. Because the singlet scattering length is one order of magnitude shorter than the triplet scattering length in all alkali-metal atoms\,\cite{bahrim2001}, its effects were initially not observed experimentally and were also neglected in the first theoretical models. Recently, it was pointed out that the hyperfine interaction of the ground-state atom leads to a mixing of singlet and triplet scattering channels in these molecules, giving rise to molecular states in which the binding results from mixed singlet and triplet e$^-$-Cs scattering\,\cite{Anderson2014b}. These states were predicted to have even weaker binding energies and were first observed in our experiments in Cs\,\cite{sassmannshausen2015b} and more recently also in Rb\,\cite{boettcher2016}. The experimental observation of Cs$_2$ long-range molecules bound by mixed singlet and triplet scattering and their theoretical modelling resulted in the first determination of a the zero-energy singlet s-wave scattering length for electron--alkali-metal-atom collisions\,\cite{sassmannshausen2015b}.

\subsection{Theoretical description}
The interaction of a ground-state atom located in the outer lobe of the Rydberg-electron orbit of another electronically highly excited atom can be reduced in good approximation to the interaction of the quasi-free Rydberg electron with the polarizable ground-state atom. This interaction, which is proportional to $\alpha/R^4$, where $\alpha$ is the polarizability of the ground-state atom and $R$ the internuclear separation, is of short range compared to the de Broglie wavelength of the Rydberg electron. Consequently, low-energy scattering theory can be applied. The electron--neutral-atom scattering wavefunctions can be expanded in partial waves labeled by the scattering angular-momentum quantum number $l$. Each partial wave is characterised in good approximation by one parameter, \emph{e.g.}, the s-wave ($l=0$) scattering length $A$ and the p-wave ($l=1$) scattering volume. At the classical turning point of the Rydberg electron, where its kinetic energy vanishes, and beyond, only s-wave scattering needs to be considered, because the centrifugal barriers of higher-$l$ partial waves render the short-range-scattering region inaccessible. Under these conditions, the interaction potential takes the simple form of a Fermi-contact pseudo-potential\,\cite{Fermi1934,Greene2000}
\begin{equation}\label{eq:Vpseudo}
V(R)=2 \pi  A(k) |\Psi(\bm{R})|^2 \;,
\end{equation}
where $A(k)$ is the energy-dependent s-wave scattering length, $k$ is the momentum of the Rydberg electron, defined by the semiclassical energy relation $-\frac{1}{n^2}=\frac{1}{2} k^2+\frac{1}{r}$ (in atomic units), and $|\Psi(\bm{R})|^2$ is the probability that the Rydberg electron is located at the position $\bm{R}$ of the neutral ground-state atom.

For the range of $n$ values we have investigated, the hyperfine interaction in the ground-state atom, the spin-orbit interaction of the Rydberg electron, and the electron-Cs scattering interaction (see Eq.\,(\ref{eq:Vpseudo})) are of similar strengths and need to be considered explicitly. The total Hamiltonian $\hat{H}$ can be written as\,\cite{Anderson2014b,sassmannshausen2015b}
\begin{equation}
\label{eq:hamiltonianI}
    \hat{H}= \hat{H}_{\rm 0} +  \hat{H}_{\rm HF} +  \hat{H}_{\rm SO} + \hat{P}_{\rm S} \cdot \hat{V}_{\rm S}+ \hat{P}_{\rm T} \cdot \hat{V}_{\rm T},
\end{equation}
where $\hat{H}_0$ is the Hamiltonian of the Rydberg atom without spin-orbit interaction, $\hat{H}_{\rm SO}=A_{\rm SO}\, \Bell_{\rm r} \cdot \bm{s}_{\rm r}$ is the spin-orbit interaction of the Rydberg electron with angular momentum $\Bell_{\rm r}$ and electron spin $\bm{s}_{\rm r}$. $\hat{H}_{\rm HF}=A_{\rm HF}\, \bm{i}_{\rm g} \cdot \bm{s}_{\rm g}$ accounts for the hyperfine interaction of the ground-state atom with nuclear spin $\bm{i}_{\rm g}$ and electron spin $\bm{s}_{\rm g}$. Singlet ($S$=0, subscript S) and triplet ($S$=1, subscript T) scattering contributions are selected using the projection operators $\hat{P}_{\rm S}=-\bm{s}_{\rm r} \cdot \bm{s}_{\rm g}+\frac{1}{4} \cdot \hat{\mathbb{1}}$ and $\hat{P}_{\rm T}=\bm{s}_{\rm r} \cdot \bm{s}_{\rm g}+\frac{3}{4} \cdot \hat{\mathbb{1}}$, respectively. We neglect molecular rotation because rotational periods of accessible states exceed the lifetimes of the molecules. We can therefore choose the molecular axis to be parallel to the $z$ axis and express the Fermi-contact-interaction operators by $\hat{V}_i = 2 \pi  A_i \hat{\delta}^3(\bm{r}-R\hat{\bm{z}})$, with $i={\rm S, T}$, for s-wave scattering between the Rydberg electron at position $\bm{r}$ and the ground-state atom located at distance $R$ along the molecular $z$-axis~\cite{Omont1977}. In a diatomic molecule, the quantum number $\Omega$ associated with the projection of the total angular momentum on the internuclear axis is a good quantum number. In our experiments, we study long-range Rydberg molecules of Cs formed by photoassociation near $n{\rm p}_{3/2} \leftarrow 6{\rm s}_{1/2}$ atomic transitions. To analyse the observations, we set up the interaction Hamiltonian\,(\ref{eq:hamiltonianI}) in the uncoupled basis $\ket{\ell_{\rm r},\lambda_{\rm r}}\ket{s_{\rm r},\sigma_{\rm r}}\ket{s_{\rm g},\sigma_{\rm g}}\ket{i_{\rm g},\omega_{i,g}}$ ($i_{\rm g}=7/2$ for $^{133}$Cs), for a single $n$p Rydberg state and the energetically closest Rydberg state, ($n-1$)d. Consequently, $\Omega= \lambda_{\rm r} + \sigma_{\rm r} + \sigma_{\rm g} + \omega_{i,g}$.

\begin{figure}[t!]
\center
\resizebox{\linewidth}{!}{%
 \includegraphics{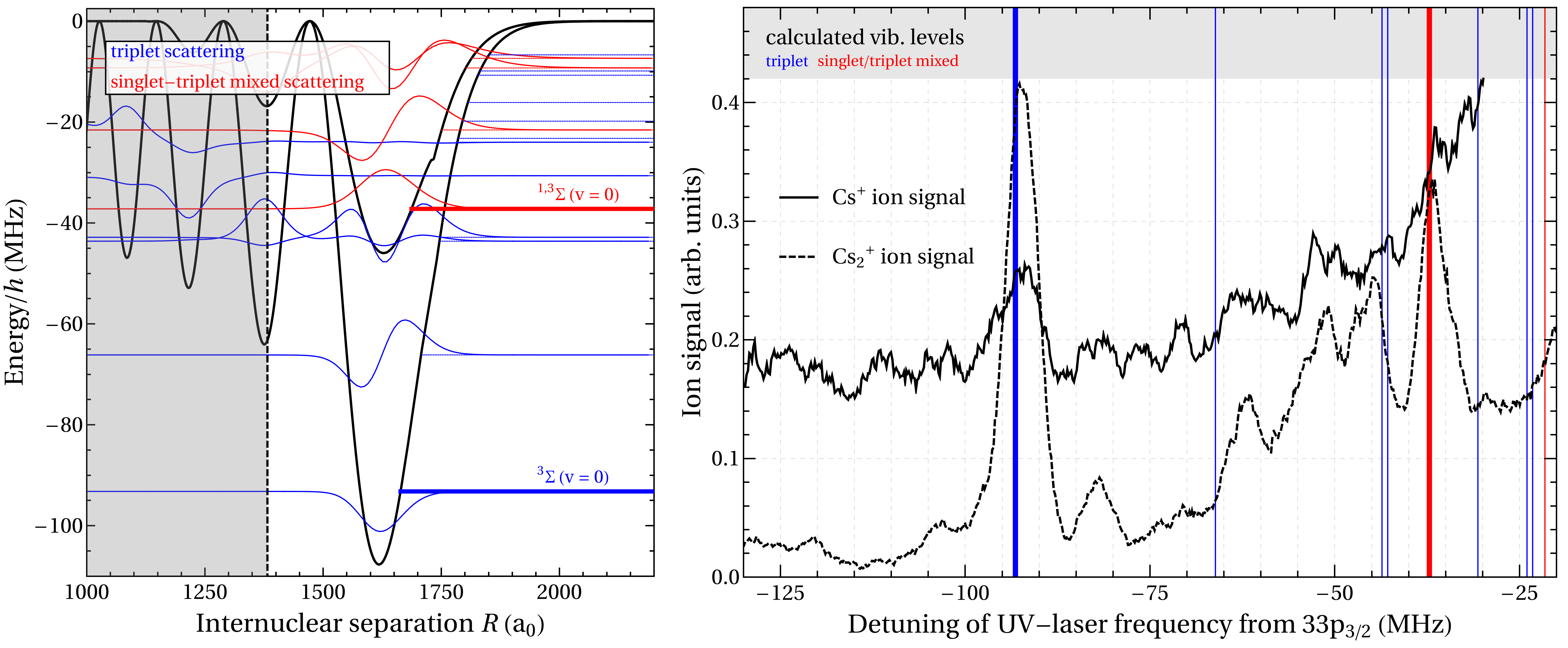} }
\caption{Calculated interaction potentials and vibrational levels (left-hand side) and photoassociation spectra recorded on the low-frequency side of the 33p$_{3/2} \leftarrow 6$s$_{1/2}, F=4$ transition of Cs (right-hand side). The singlet/triplet mixed states are shown in red and the purely triplet scattering states are shown in blue. The vertical dashed line in the panel on the left-hand side denotes the distance where the semiclassical momentum of the Rydberg electron corresponds to the lowest P$_{3}$ resonance in the e-Cs scattering. The figure is adapted from\,\cite{chimia2016}.}
\label{fig:3}
\end{figure}

In our model\,\cite{sassmannshausen2015b}, we restrict the interaction of the Rydberg electron and the ground-state Cs atom to s-wave scattering, so that the binding is proportional to ${|\Psi(z=R)|}^2$. With our choice of coordinates, only the $\lambda_{\rm r}=0$ orbital has a nonzero electron density at the position of the ground state atom. Therefore, only the $\lambda_{\rm r}=0$ component of the Rydberg-electron wavefunction contributes to the binding. Levels associated with different values of $\lambda_i$ are, however, mixed by the spin-orbit interaction, and only states with $\omega_{\rm r}=\lambda_{\rm r}+\sigma_{\rm r} = \sigma_{\rm r}$ contribute to the molecular binding. Because of this restriction, the spin-orbit interaction does not mix singlet and triplet scattering channels. No such restriction exists for the hyperfine interaction of the ground-state atom, which couples the electron spin $\bm{s}_{\rm g}$ and the nuclear spin $\bm{i}_{\rm g}$ to form $\bm{F} = \bm{i}_{\rm g} + \bm{s}_{\rm g}$. Because of the hyperfine interaction of the ground-state atom, the quantum number $\sigma_{\rm g}$ of the projection of the electron spin of the ground-state atom is not a good quantum number, which results in mixing of singlet and triplet scattering channels. The eigenstates and eigenvalues of the Hamiltionan defined in Eq.\,(\ref{eq:hamiltonianI}) form two sets of molecular states. In the first set, to which we assign the label $^{3}{\rm \Sigma}$ (this label only refers to the scattering and not to the overall molecular symmetry), the binding results from pure triplet scattering channels. In the second set, which we designate with the label $^{1,3}{\rm \Sigma}$, the binding results from mixed singlet and triplet scattering channels. The states resulting from pure triplet scattering are more deeply bound because in all alkali-metal atoms the singlet scattering lengths are smaller than the triplet scattering lengths. A pure singlet scattering molecular state does not exist. Exemplary potential curves calculated for the ${\rm ^3\Sigma}$ and $^{1,3}{\rm \Sigma}$ state correlated to the 33p$_{3/2}, 6$s$_{1/2}, F=4$ dissociation asymptote are depicted on the left-hand side of Fig.\,\ref{fig:3}. We find that the binding energies of pure triplet s-wave scattering states are independent of the hyperfine state of the ground-state atom, whereas the binding energies of the singlet-triplet mixed state depend on the hyperfine state of the ground-state atom ($F=3,4$). B\"ottcher \textit{et al.} found that all pure triplet states except the state with maximum $\Omega$ acquire singlet character when an external magnetic field is applied\,\cite{boettcher2016}.

Our s-wave scattering model is the simplest model that reproduces the positions of the vibrational ground states in the outer-most well of the potentials of Cs$_2$ (see Fig.\,\ref{fig:3}). To explain vibrationally excited levels and further resonances observed on the low and high-frequency side of atomic $n$p$_{3/2}$ Rydberg states, the contributions from higher partial waves in the electron--ground-state-atom scattering must be considered, most importantly p-waves (see, \emph{e.g.}, Refs.\,\cite{bendkowsky2010,butscher2011}). The binding induced by p-wave scattering is proportional to the gradient of the electronic wavefunction. While molecular states of the same $\Omega$ values are degenerate in a pure s-wave scattering model, their degeneracy is lifted by p-wave-scattering contributions.

\subsection{Experimental}
\label{experimental}
All experiments are performed with ultracold samples of Cs atoms released from a vapour-loaded compressed magneto-optical trap (MOT) at a temperature of 40\,$\upmu$K and a density of $10^{11}$\,cm$^{-3}$. Temperature and density are estimated on the basis of saturated-absorption images of the ultracold cloud. We use single-photon $n$p$_{3/2} \leftarrow 6$s$_{1/2}$ transitions to excite p$_{3/2}$ Rydberg states of Cs directly from the electronic ground state. A Cooper minimum in the photoexcitation cross section of the $\epsilon {\rm p}_{1/2} \leftarrow 6$s$_{1/2}$ channel just above the ionisation threshold\,\cite{raimond1978} strongly suppresses the excitation of p$_{1/2}$ Rydberg states, to the extent that $n{\rm p}_{1/2} \leftarrow 6{\rm s}_{1/2}$ transitions at high $n$ values are many orders of magnitude weaker than $n$p$_{3/2} \leftarrow 6$s$_{1/2}$ transitions and thus not observed. The UV radiation with a wavelength of about $319$\,nm necessary to excite high Rydberg states of Cs from the 6s$_{1/2}$ ground state is generated with a single-mode cw ring dye laser, the output of which is frequency doubled in a BBO crystal inside an external enhancement cavity. The fundamental frequency of the ring laser is locked to its internal reference cavity, resulting in a bandwith of $\sim 3$\,MHz in the UV. The absolute frequency is calibrated with a wavemeter (HighFinesse, WS-7) and relative frequencies with respect to a frequency-stabilised HeNe laser (Thorlabs, 2 mW, stabilised to 2 MHz) are measured with a scanning cavity. The final calibration accuracy of frequency detunings from the strong atomic transitions is $\sim5$\,MHz. For the results presented in this section, 40\,$\upmu$s long UV pulses are produced using an electro-optical modulator and the UV-laser beam is focused to a spot size of $\sim150$\,$\upmu$m in the centre of the ultracold Cs atom cloud. All Rydberg-excitation experiments are performed once with all atoms prepared in the $F=3$ hyperfine component of the 6s$_{1/2}$ electronic ground state and once with all atoms pumped to the upper $F=4$ component. Rydberg atoms and long-range molecules are detected as atomic and molecular ions, respectively, on a microchannel-plate (MCP) detector following spontaneous or pulsed field ionisation. Cs$^+$ and Cs$_2^+$ ions are detected separately by setting temporal detection gates at the corresponding positions of the ion-time-of-flight spectrum. During the photoassociation experiments, all trapping fields are switched off and stray magnetic and electric fields are suppressed to below 20\,mG and 20\,mV/cm, respectively. Further technical details concerning the experimental setup can be found in Refs.\,\cite{sassmannshausen2013,sassmannshausen2015b}.

\subsection{Results}
\begin{figure}[t!]
\center
\resizebox{0.49\linewidth}{!}{%
 \includegraphics{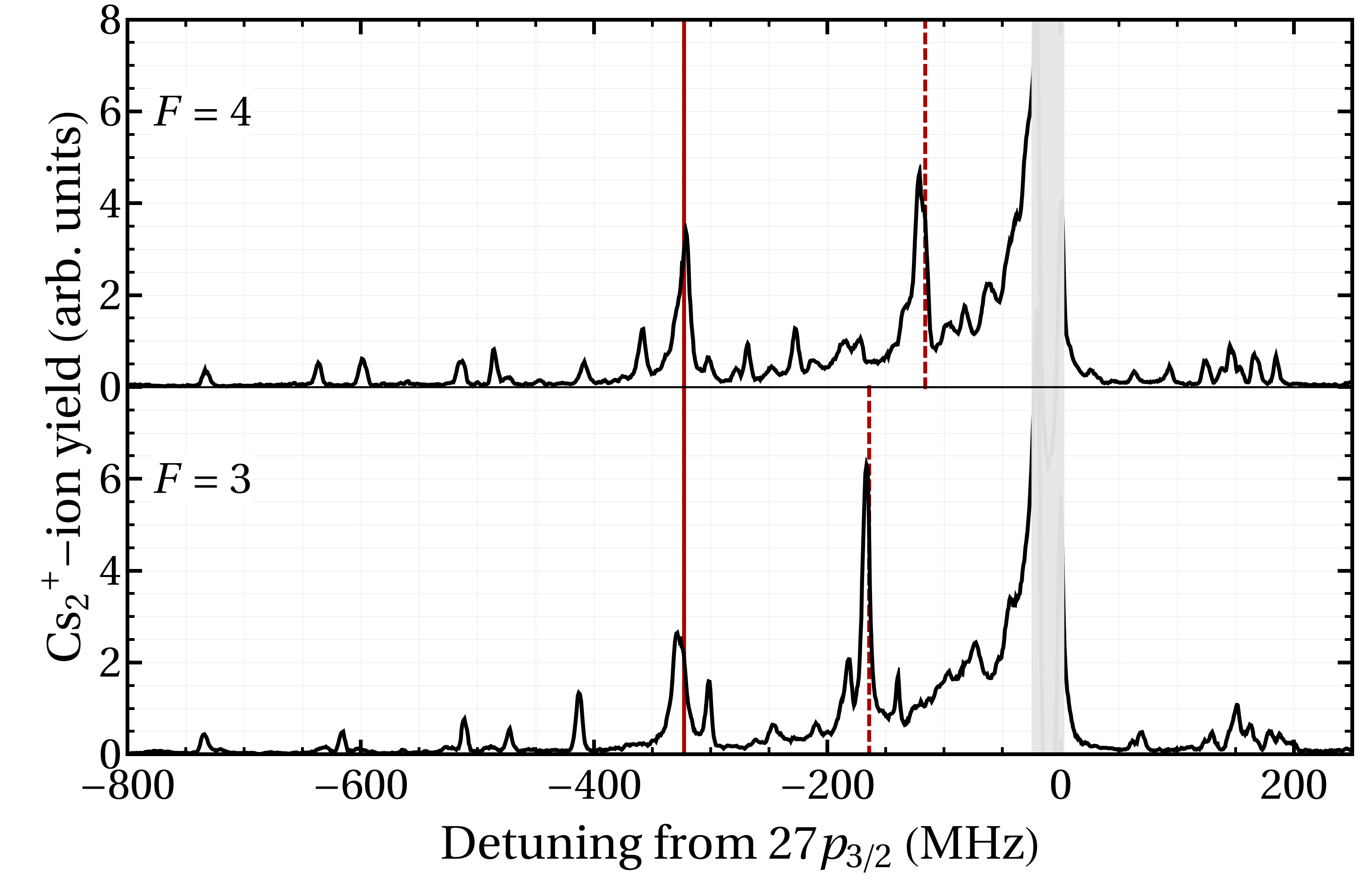}}
\resizebox{0.49\linewidth}{!}{%
 \includegraphics{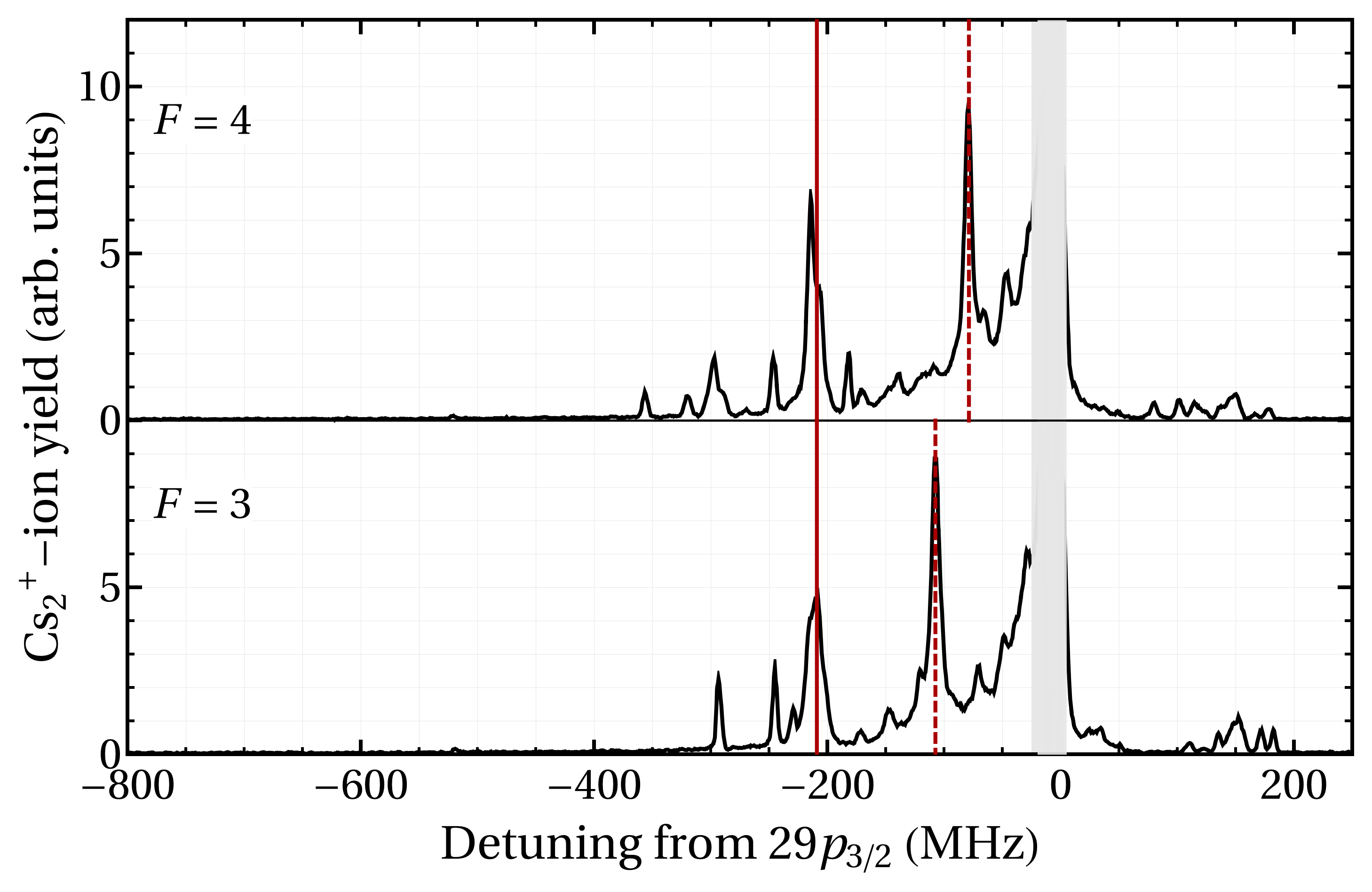}}
\caption{UV-laser spectra in the vicinity of $n$p$_{3/2} \leftarrow 6$s$_{1/2}$ transitions of Cs with $n=27$ (left panel) and $n=29$ (right panel). The spectra represent the spontaneously produced Cs$^+_2$-ion signal obtained after application of a 40-$\upmu$s-long UV pulse. For both values of the principal quantum number, the two spectra were recorded after preparing the atoms either in the $F=4$ hyperfine level (upper traces) or in the $F=3$ hyperfine state (lower traces) of the 6s$_{1/2}$ electronic ground state. The two most intense photoassociation resonances in each spectrum correspond closely to the predicted positions of the vibrational ground states of the pure-triplet-scattering potentials (solid red lines) and of the mixed-singlet- and triplet-scattering potentials (dashed red lines).}
\label{fig:4}
\end{figure}

We have carried out measurements in the range of principal quantum number $n=26-34$. Exemplary spectra at $n=27$ and $n=29$ are presented in Fig.\,\ref{fig:4} and reveal sharp photoassociation lines at positive and negative detunings from the $n$p$_{3/2} \leftarrow 6$s$_{1/2}$ transition frequencies. At the positions of the strongest resonances, less than 50 ions are detected per experimental cycle, which corresponds to a density of long-range Rydberg molecules of 10$^6$\,cm$^{-3}$. Under these conditions, the signals of spontaneously produced Cs$^+$ ions are weak and mainly Cs$_2^+$ ions are detected. The positions of some molecular resonances depend on the hyperfine component $F$ in which the ground-state atoms are prepared, whereas the positions of other resonances do not depend on the selected hyperfine component of the ground state. The two most intense lines in each spectrum (marked by dashed red lines in Fig.\,\ref{fig:4}) behave differently in this respect: The positions of the molecular states observed at larger detunings from the dissociation asymptotes (given by the $n$p$_{3/2} \leftarrow 6$s$_{1/2}$ atomic resonance frequency) are independent of $F$, whereas the less deeply bound molecular states are observed at larger detunings for $F=3$ than for $F=4$. These intense lines are assigned to the vibrational ground states of the two potentials predicted by our calculations (see Fig.\,\ref{fig:3}). The more strongly bound molecular state results from triplet s-wave scattering, whereas the less strongly bound state results from a mixing of singlet and triplet s-wave scattering channels. The vertical dashed red lines in Fig.\,\ref{fig:4} indicate the calculated positions of these levels based on the model presented in Section\,2.2, after adjustment of the zero-energy s-wave singlet- and triplet-scattering lengths to $A_{\rm 0,S}=(-3.5 \pm 0.4) a_0$ and $A_{\rm 0,T}=(-21.8 \pm 0.2)a_0$, respectively\,\cite{sassmannshausen2015b}.

The experimentally determined binding energies of the vibrational ground states of the long-range Cs$_2$ molecules correlated to Cs\,$n$p$_{3/2}$ + Cs\,6s$_{1/2}$ asymptotes are compared to the predictions of our s-wave scattering model with adjusted zero-energy scattering lengths in Table\,\ref{tab:f3} for $F=3$ and Table\,\ref{tab:f4} for $F=4$. All 36 resonances are reproduced by the calculations within better than 10\,MHz. The $^{1,3}{\rm \Sigma}$ levels correlated to the 32p$_{3/2}$\,6s$_{1/2}$\,$F=3,4$ asymptotes represent an exception to the trend of decreasing binding energies with increasing $n$ value. This anomaly is caused by the accidental near-degeneracy of the 32p$_{3/2}, 6$s$_{1/2} (F=3)$ level with the 32p$_{1/2}, 6$s$_{1/2} (F=4)$ level which arises because the spin-orbit splitting of the 32p state almost exactly matches the hyperfine-structure splitting of the 6s$_{1/2}$ ground state. The effect of this perturbation is also reflected in the values of the calculated molecular dipole moments listed in the ninth column of Table\,\ref{tab:f3}.

These electric dipole moments are calculated in the frame of the non-rotating molecule and arise in our model from the mixing of atomic wavefunctions with different orbital angular momentum induced by the contact-interaction Hamiltonian Eq.\,(\ref{eq:Vpseudo}). This interaction reduces the spherical symmetry of the Rydberg-atom Hamiltonian to cylindrical symmetry and is in that respect similar to the Stark effect for atoms in homogeneous electric fields. In the laboratory frame, the near degeneracy of rotational levels of the long-range Rydberg molecules leads to the onset of a linear Stark effect already at very low field strengths below 1\,mV/cm. Our calculations of the molecular dipole moment account for the observed linear Stark effect in long-range Rydberg molecules (see also\,\cite{li2011,sassmannshausen2015b,booth2015}), which results from the increased Rydberg-electron density close to the ground-state atom. Our calculations predict the existence of molecular states with ``negative'' dipole moments, corresponding to a reduced Rydberg electron density in the vicinity of the ground-state atom. This situation differs from the behaviour usually encountered in long-range Rydberg molecules\,\cite{li2011,booth2015}.

The assignment of vibrationally excited states of the ${\rm ^3\Sigma}$ and $^{1,3}{\rm \Sigma}$ potentials and of molecular resonances observed both at detunings below the vibrational ground state of the $^3{\rm \Sigma}$ potential and on the high-frequency side of the atomic $n$p$_{3/2} \leftarrow 6$s$_{1/2}$ transitions is beyond the scope of the s-wave scattering model presented above. The influence of p-wave scattering on the energy-level structure of long-range Rydberg molecules of the alkali-metal atoms, which have a low-energy $^3$P shape resonance in electron-atom scattering\,\cite{scheer1998}, is significant already at small displacements from the classical outer turning point of the Rydberg electron\,\cite{bendkowsky2010}. The divergence of the $A_{^3{\rm P}}$ scattering volume at the position of the shape resonance leads to the formation of an unusual molecular state, which has been referred to as the ``butterfly'' state\,\cite{hamilton2002}. The Cs\,(6s$_{1/2}$)\,+\,Cs$^*$ dissociation asymptote of this state corresponds to the position of the hydrogenic high-$l$ manifold and its binding energy corresponds to several hundred GHz. Consequently, the  potential of the ``butterfly'' state undergoes avoided level crossings with lower-lying molecular states correlated to Cs\,(6s$_{1/2}$)\,+Cs\,($nl_J$) dissociation asymptotes with $l=0,1,2,3$. Resonances on the high-frequency side of the atomic resonance could originate from the upper branches of the avoided curve crossings and are thus similar to the long-range molecular states reported on the high-frequency side of 6s$_{1/2}$+$n$s$_{1/2}$ asymptotes in Cs\,\cite{Tallant2012}. Regular weak series of molecular states observed at large negative detunings in Fig.\,\ref{fig:4} are likely to result from almost harmonic potential wells of the ``butterfly'' potential. Experiments on the Stark shifts and broadenings of these levels\,\cite{sass_unpublished} indicate larger dipole moments than observed here for the vibrational ground states in the outer wells of s-wave-scattering potentials, in accordance with the high-$l$-mixed character of a ``butterfly-like'' molecular state. We are currently setting up an extended potential model including the p-wave scattering contributions, with which we hope to assign all remaining resonances in our experimental photoassociation spectra.

\begin{table}
\centering
\caption{Measured and calculated binding energies of ${\rm Cs}(n{\rm p}_{3/2})-{\rm Cs}(6{\rm s}_{1/2},F=3)$ long-range Rydberg molecules, $E^{\rm expt}/h$ and $E^{\rm theor}/h$, respectively, in units of MHz. Average dipole moments $\bar{\mu}^{\rm theor}$ of the corresponding molecular states of different $\Omega$ (quantum number of the projection of the total angular momentum onto the molecular axis), which are degenerate in zero electric field, are given in units of Debye and have been calculated at the predicted equilibrium distances $R^{\rm eq}$ (values given in units of nm).}
\label{tab:f3}
\begin{tabular}{c|cccccccc}
\hline\hline\noalign{\smallskip}
$n$&$E_{\rm ^3\Sigma}^{\rm expt}$&$E_{\rm ^3\Sigma}^{\rm theor}$&$R_{\rm ^3\Sigma}^{\rm eq}$&$\bar{\mu}_{\rm ^3\Sigma}^{\rm theor}$&$E_{\rm ^{1,3}\Sigma}^{\rm expt}$&$E_{\rm ^{1,3}\Sigma}^{\rm theor}$&$R_{\rm ^{1,3}\Sigma}^{\rm eq}$&$\bar{\mu}_{\rm ^{1,3}\Sigma}^{\rm theor}$\\
\noalign{\smallskip}\hline\noalign{\smallskip}
 26 & $-$406 & $-$409 & 46.8 & 19.2 & $-$209 & $-$205 & 47.1 & 20.5 \\
 27 & $-$327 & $-$324 & 51.1 & 19.0 & $-$167 & $-$164 & 51.4 & 20.6 \\
 28 & $-$251 & $-$259 & 55.6 & 18.9 & $-$127 & $-$132 & 55.9 & 20.5 \\
 29 & $-$211 & $-$208 & 60.3 & 18.7 & $-$107 & $-$106 & 60.6 & 20.0 \\
 30 & $-$172 & $-$169 & 65.2 & 18.5 & $-$88 & $-$86 & 65.5 & 19.3 \\
 31 & $-$137 & $-$138 & 70.4 & 18.3 & $-$68 & $-$70 & 70.6 & 17.6 \\
 32 & $-$118 & $-$113 & 75.7 & 16.6 & $-$43 & $-$45 & 76.3 & 4.5 \\
 33 & $-$90 & $-$93 & 81.2 & 17.7 & $-$51 & $-$51 & 81.3 & 23.8 \\
 34 & $-$74 & $-$77 & 86.9 & 17.5 & $-$38 & $-$42 & 87.1 & 22.0 \\
\noalign{\smallskip}\hline
\end{tabular}
\end{table}

\begin{table}
\centering
\caption{Measured and calculated binding energies of ${\rm Cs}(n{\rm p}_{3/2})-{\rm Cs}(6{\rm s}_{1/2},F=4)$ long-range Rydberg molecules $E^{\rm expt}/h$ and $E^{\rm theor}/h$, respectively, in units of MHz. Average dipole moments $\bar{\mu}^{\rm theor}$ of the corresponding molecular states of different $\Omega$ (quantum number of the projection of the total angular momentum onto the molecular axis), which are degenerate in zero electric field, are given in units of Debye and have been calculated at the predicted equilibrium distances $R^{\rm eq}$ (values given in units of nm).}
\label{tab:f4}
\begin{tabular}{c|cccccccc}
\hline\hline\noalign{\smallskip}
$n$&$E_{\rm ^3\Sigma}^{\rm expt}$&$E_{\rm ^3\Sigma}^{\rm theor}$&$R_{\rm ^3\Sigma}^{\rm eq}$&$\bar{\mu}_{\rm ^3\Sigma}^{\rm theor}$&$E_{\rm ^{1,3}\Sigma}^{\rm expt}$&$E_{\rm ^{1,3}\Sigma}^{\rm theor}$&$R_{\rm ^{1,3}\Sigma}^{\rm eq}$&$\bar{\mu}_{\rm ^{1,3}\Sigma}^{\rm theor}$\\
\noalign{\smallskip}\hline\noalign{\smallskip}
 26 & $-$409 & $-$409 & 46.7 & 26.3 & $-$144 & $-$141& 47.5 & 5.4 \\
 27 & $-$323 & $-$324 & 51.0 & 26.2 & $-$121 & $-$115 & 51.7 & 5.5 \\
 28 & $-$250 & $-$259 & 55.5 & 26.2 & $-$89 & $-$94 & 56.2 & 5.7 \\
 29 & $-$212 & $-$208 & 60.2 & 25.9 & $-$79 & $-$78 & 60.9 & 5.9 \\
 30 & $-$170 & $-$169 & 65.1 & 25.7 & $-$65 & $-$64 & 65.7 & 6.1 \\
 31 & $-$141 & $-$138 & 70.3 & 25.5 & $-$53 & $-$53 & 70.8 & 6.3 \\
 32 & $-$116 & $-$113 & 75.5 & 25.1& $-$46 & $-$44 & 76.1 & 6.4 \\
 33 & $-$89 & $-$93 & 81.2 & 24.9 & $-$35 & $-$37 & 81.6 & 6.7 \\
 34 & $-$74 & $-$77 & 86.8 & 24.6 & $-$31 & $-$31 & 87.3 & 7.0 \\
\noalign{\smallskip}\hline
\end{tabular}
\end{table}
\section{Interacting Rydberg-atom-pair states}
The prospect of realising a quantum gate\,\cite{jaksch2000} on the basis of the Rydberg excitation blockade\,\cite{lukin2001} has been the motivation for many recent experiments with ultracold Rydberg atoms, see \emph{e.g.}, Refs.\,\cite{urban2009,Gaetan2009,schwarzkopf2011,tong2003,Schauss2012,barredo2014,hankin2014}. The majority of the experimental observations can be understood in terms of a simple $C_6/R^6$ scaling of the long-range van der Waals interaction between two Rydberg atoms and the concept of an excitation-blockade radius
\begin{equation}
\label{eq:blockaderadius}
R_{\rm B} = \sqrt[6]{2C_6/(\delta\nu_{\rm L}h)}.
\end{equation}
$R_{\rm B}$ corresponds to the minimal distance at which two ground-state atoms can both be excited to a high-$n$ Rydberg state by resonant excitation with a laser-excitation bandwidth  $\delta\nu_{\rm L}$. In Eq.\,\ref{eq:blockaderadius}, $C_6$ is the van der Waals coefficient of the Rydberg-Rydberg interaction, which scales as $n^{11}$. However, the simple treatment of the Rydberg-excitation blockade using van der Waals coefficients fails at short internuclear distances, where the interactions between the Rydberg atoms become large and the system cannot accurately be described using a perturbative treatment of the long-range interactions. Our experiments focus on this regime where a complex molecular-level structure emerges as a result of the very high state density and multiple (avoided) level crossings.

To reach this regime, we overcome the excitation blockade using high-intensity pulses\,\cite{Demekhin2013} with a bandwidth of 130\,MHz or a resonant two-colour two-photon excitation scheme. The resulting Rydberg-atom-pair states, also referred to as Rydberg macrodimers, have been observed previously at high $n$ values around $n=70$ in Rb\,\cite{farooqi2003,stanojevic2006,stanojevic2008} and Cs\,\cite{overstreet2007,overstreet2009}. In our experiments, we investigate these states at lower $n$ values, which offer the advantages of a lower state density, of simpler assignments and of model calculations necessitating smaller basis sets. In the UV-laser spectra presented here, the excitation of the Rydberg-atom-pair states leads to additional sharp lines close to $n$p$_{3/2} \leftarrow 6$s$_{1/2}$ resonances in the range $n=22-36$. Our experiments are not only of relevance in the context of the excitation blockade mentioned above, but they also reveal that the same interactions can give rise to weakly bound diatomic molecules with internuclear separations in the micrometer range\,\cite{overstreet2009}.

We have developed a long-range potential model (described in Section\,\ref{sec:theo}) to account for the interactions between neighbouring Rydberg atoms and interpret the molecular resonances as resulting from these interactions. The model is based on a multipole expansion of the long-range interaction Hamiltonian in an uncoupled atom-pair-state basis and the determination of its eigenvalues and eigenvectors. With this model, we calculate the excitation probability as a function of the internuclear separation of the two Cs atoms and in this way derive an ``exact'' excitation blockade radius (see lower panels in Fig.\,\ref{fig:5}). We simulate our experimental spectra by calculating the excitation probability as a function of the laser detuning, summing over all molecular states and integrating over all internuclear distances and all orientations of the internuclear axis in space.

The experimental results obtained with a 130-MHz-bandwidth intense pulsed UV laser are presented in Section\,\ref{sec:pulsed}. In Section\,\ref{sec:seed}, we introduce an alternative two-colour excitation scheme, which relies on the resonant excitation of ``seed'' Rydberg atoms and discuss its advantages for the study of Rydberg macrodimers.

\subsection{Long-range potential model}
\label{sec:theo}
\begin{figure}[t!]
\center
\resizebox{0.99\linewidth}{!}{%
 \includegraphics{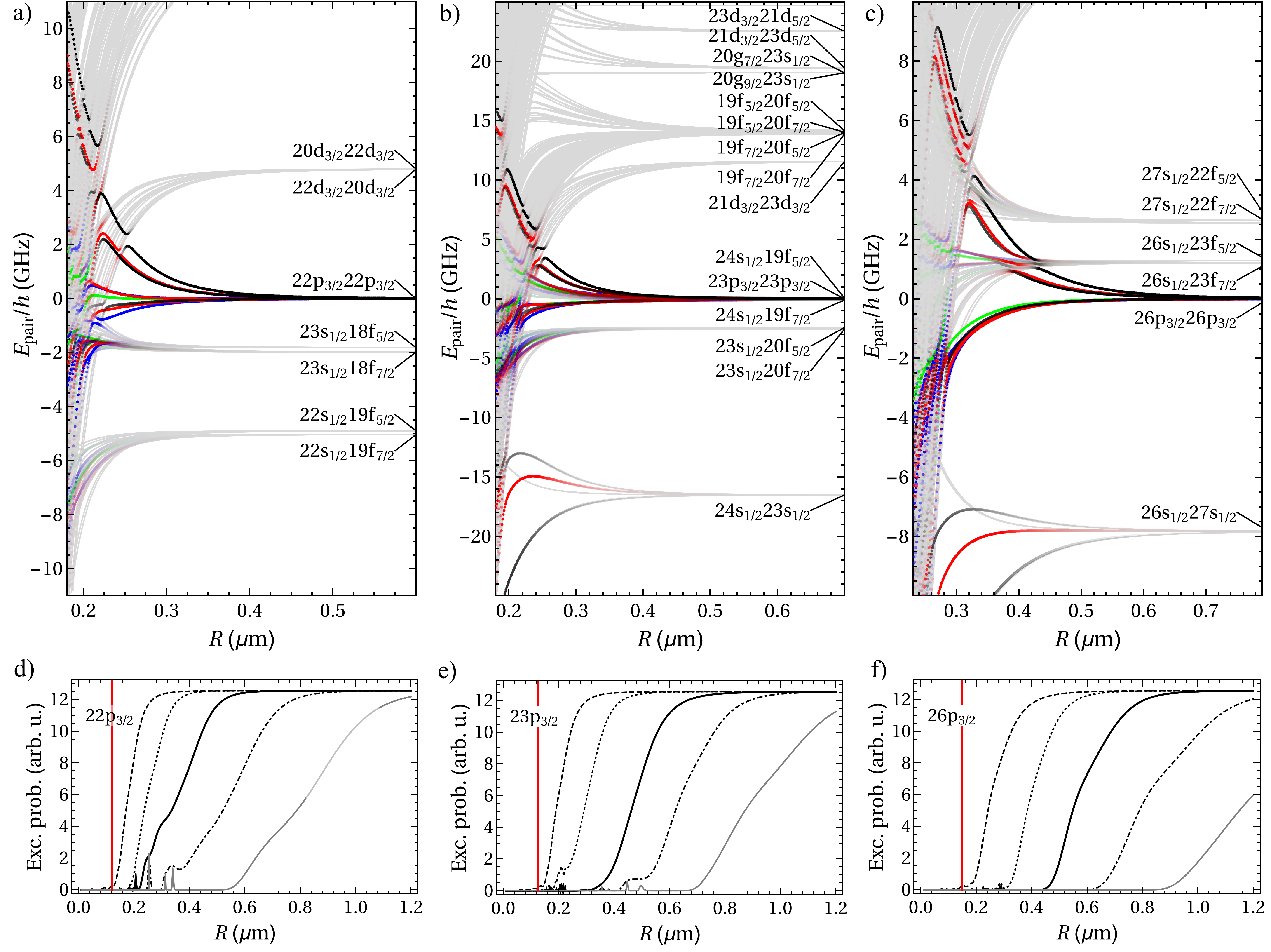}}
\caption{Potential-energy curves of long-range interacting Rydberg-atom-pair states in the vicinity of the 22p$_{3/2}$22p$_{3/2}$ (a), 23p$_{3/2}$23p$_{3/2}$ (b), and 26p$_{3/2}$26p$_{3/2}$ (c) dissociation asymptote. The colour shading of the potential curves represents the value of $\overline{p_\Psi}(R)$ as defined in the text. Full colour denotes 5\% $n$p$_{3/2}n$p$_{3/2}$ character. Black, red, blue and green curves correspond to molecular states with $\Omega=0,1,2,3$, respectively. Panels d) to f) show the corresponding probability of resonant $n$p$_{3/2} \leftarrow 6$s$_{1/2}$ transitions as a function of the internuclear distance to another Cs atom in the $n$p$_{3/2}$ state for different laser bandwidths of 10\,GHz (dashed trace), 1\,GHz (dotted trace), 100\,MHz (thick solid trace), 10\,MHz (dotted-dashed trace), and 1\,MHz (solid grey trace). The LeRoy radii for the $n$p$_{3/2}n$p$_{3/2}$ Rydberg-atom pairs are indicated by vertical red lines.}
\label{fig:5}
\end{figure}

Many of the observed spectral structures originate from avoided crossings between different van der Waals potentials and from the interactions of many different pair states. Consequently, a simple treatment of interaction-induced shifts with a single long-range van der Waals coefficient (as in Eq.\,\ref{eq:blockaderadius}) is not adequate. We found it necessary to perform a full treatment of the interaction Hamiltonian, which we set up using an uncoupled atom-pair-state basis $\ket{ n_{\rm A}l_{\rm A}j_{\rm A}\omega_{\rm A}, n_{\rm B}l_{\rm B}j_{\rm B}\omega_{\rm B}}$\,\cite{sassmannshausen2015}. In this basis $l_i$, $j_i$ and $\omega_i$ ($i={\rm A, B}$) represent the quantum numbers of the Rydberg-electron orbital angular momentum, of the total electronic angular momentum and its projection onto the interatomic axis, respectively. This notation can be shortened to $\ket{\gamma_\AR  j_\AR \omega_\AR, \gamma_\BR j_\BR \omega_\BR}$. Because all effects observed in our experiments are caused by interactions at distances greater than the LeRoy radius  $R_\textrm{LR}=2\left(\braket{r_A^2}^{1/2}+\braket{r_B^2}^{1/2}\right)$~\cite{leroy1973}, the multipole expansion\,\cite{flannery2005}
\begin{align}\label{eq:vint}
V_\textrm{inter}(R) & =  \sum_{\mathclap{L_{\AR/\BR}=1}\hspace{0.5cm}}^\infty \sum_{\hspace{0.5cm}\mathclap{\omega=-L_<}}^{+L_<}\frac{(-1)^{L_\BR}f_{L_\AR,L_\BR,\omega}}{R^{L_\AR+L_\BR+1}} Q_{L_\AR,\omega}(\vec{r}_\AR)Q_{L_\BR,-\omega}(\vec{r}_\BR),  \\
\intertext{with} Q_{L,\omega}(\vec{r}) &= \sqrt{\frac{4 \pi}{2 L +1}} r^L Y_{L,\omega}(\hat{r}) \label{eq:multipoles} \\
\intertext {and}  f_{L_\AR,L_\BR,\omega} & =  \frac{(L_\AR+L_\BR)!}{\sqrt{(L_\AR+\omega)!(L_\AR-\omega)!(L_\BR+\omega)!(L_\BR-\omega)!}}
\end{align}
is well suited to describe the long-range interactions between two Cs Rydberg atoms (labeled A and B). Terms up to the octupole-octupole contribution were taken into account. The total Hamiltonian is given by $H =H^\AR_0+H^\BR_0+V_\textrm{inter}(R)$, where $H^i_0$ are the Hamiltonians of the unperturbed Rydberg atoms including spin-orbit interaction and $i={\rm A,B}$ (see Eq.\,\ref{eq:intro}). The eigenenergies $E_\Psi(R)$ of molecular states $\Psi$ are determined as a function of the internuclear distance. The molecular eigenfunctions are linear combinations of the basis functions, \emph{i.e.}, $\ket{\Psi(R)}=\sum c_{\gamma_\textrm{A} j_\textrm{A}  \omega_\textrm{A}, \gamma_\textrm{B} j_\textrm{B} \omega_\textrm{B}}(R)\ket{\gamma_\textrm{A}  j_\textrm{A} \omega_\textrm{A}, \gamma_\textrm{B} j_\textrm{B} \omega_\textrm{B}}$. Exemplary potential-energy curves are presented in the upper panels of Fig.\,\ref{fig:5} for interacting Rydberg-atom-pair states of Cs in the vicinity of $n$p$_{3/2} + n$p$_{3/2}$ dissociation asymptotes with $n=22,23,26$. Convergence of the calculations was reached for basis sets with $\mathrm{\Delta} n_{\rm max}^* =1$, $l \leq 6$ and an energy range of $0.64 \cdot 4 R_{\rm Cs}/(n^{*})^3$ centred around a given $n$p$_{3/2} + n$p$_{3/2}$ asymptote ($n^*=n-\delta_{n,l}$ is the effective principal quantum number and $R_{\rm Cs}$ is the Rydberg constant of the Cs atom). The magnitude of the matrix elements of $V_{\rm inter}$ decreases rapidly with the difference in the effective principal quantum numbers of the two interacting Rydberg atoms so that only the coupling of atoms with $\mathrm{\Delta} n^* \leq 1$ needs to be considered. Even with this truncated basis, the density of Rydberg-atom-pair states at high $n$ values and short internuclear distances is enormous. It is therefore crucial to identify the observable states, which are those having significant contributions from the optically accessible $n$p$_{3/2}n$p$_{3/2}$ states. To calculate these contributions, the molecular states must be rotated from the molecular axis system to the laboratory-fixed axis system using Wigner $d_{\omega m}^{j}(\theta)$ rotation matrices\,\cite{walker2008}. We determine the $n$p$_{3/2}n$p$_{3/2}$ character of a molecular state $\Psi$ as a function of the internuclear separation $R$ and the polar angle $\theta$ between the $z$ axis of the laboratory system and the molecule-fixed $z$ axis (there is no dependence on the azimuthal angle) by evaluating
\begin{align}
p_\Psi(\theta,R)&=\left|\braket{\Psi(R)|\phi}\right|^2= \nonumber \\
&\Big|\sum_{\mathclap{\substack{\gamma_\AR j_\AR  \omega_\AR, \gamma_\BR j_\BR \omega_\BR  \\ j_\AR m_\AR, j_\BR m_\BR}}}\braket{\Psi(R)|\gamma_\AR j_\AR  \omega_\AR, \gamma_\BR j_\BR \omega_\BR } d_{\omega_{\AR} m_{\AR}}^{j_{\AR}}(\theta) \times \nonumber
\\ & \hspace{1.5cm} d_{\omega_{\BR} m_{\BR}}^{j_{\BR}}(\theta) \braket{j_\AR m_\AR, j_\BR m_\BR|\phi}\Big|^2, \hspace{0.2cm}
\label{eq:rot}
\end{align}
where $\ket{\phi}$ are the optically accessible basis states (for $n$p$_{3/2}n$p$_{3/2}$ we take $l=1$,$j_{\rm A}=j_{\rm B}=3/2$) which are conveniently expressed in the laboratory-fixed frame. To quantify signal strengths expected for the photoassociation process of two atoms in an ultracold gas to diatomic molecules with randomly oriented internuclear axes, $p_\Psi(\theta,R)$ is averaged over all angles and we obtain $\overline{p_\Psi}(R)=\int p_\Psi(\theta,R)2 \pi {\rm sin}(\theta) \mathrm{d} \theta$. The quantity $\overline{p_\Psi}(R)$ is used to indicate the spectral intensity of the molecular potentials in Fig.\,\ref{fig:5} by means of the colour shading.

The full theoretical line profile for the two-photon excitation of interacting Rydberg-atom-pair states is calculated as
\begin{align}
\label{eq:lineprofile}
    s(E) =16 \pi \sum_\Psi \int_0^\infty \frac{\omega_\textrm{atom}^4}{(E_{\rm pp}-E_{\Psi}(R))^2} G(E-E_\Psi(R)) \overline{p_\Psi}(R) R^2 \textrm{d}R.
\end{align}
In Eq.\,\ref{eq:lineprofile}, $G(E)$ is the experimental laser line-shape function and $\omega_\textrm{atom}$ is the atomic Rabi frequency of the relevant $n$p$_{3/2} \leftarrow 6$s$_{1/2}$ transition. The atomic Rabi frequency is not calculated but used as a global scaling factor to reproduce experimental spectra. The simulated line profiles depicted in Figs.\,\ref{fig:1},\,\ref{fig:6}, and\,\ref{fig:7} as solid traces are calculated using Eq.\,\ref{eq:lineprofile}.

On the basis of the long-range potential curves, the two-atom excitation probability for resonant Rydberg excitation can be determined as a function of the internuclear separation (see lower panels of Fig.\,\ref{fig:5}). The region where this excitation probability approaches zero corresponds to the effective blockade radius, which is often approximated by Eq.\,\ref{eq:blockaderadius}. In the experiments described here, the two-atom-excitation probability only reaches zero at distances significantly smaller than the mean nearest-neighbour separation in the ultracold gas, both for the pulsed excitation and the cw excitation with 130\,MHz and 1\,MHz excitation bandwidths, respectively. Consequently, no significant blockade effect is expected in the range of $n$ values measured here\,\footnote{This theoretical prediction is confirmed experimentally. Both with pulsed and narrow-band cw excitation the atom cloud can be completely depleted when the laser frequency is set to an atomic $n$p$_{3/2}$ resonance (with $22<n<34$) and the laser power is gradually increased.}. The reason for the small values of $R_{\rm B}$ encountered here in $n$p$_{3/2}$ Rydberg states are molecular states that experience almost zero level shift even at short internuclear distances. These states have been termed ``F\"orster zero'' states and appear in Rydberg states with $l>0$\,\cite{walker2005,walker2008}. Our calculations show that the negative effects of these ``F\"orster zero'' states on blockade experiments are avoided by using excitation to $n$s Rydberg levels, and indeed, the most successful blockade experiments reported to date rely on s Rydberg states\,\cite{urban2009,Gaetan2009}. However, Rydberg-blockade effects have also been demonstrated in p\,\cite{hankin2014} and d\,\cite{barredo2014} Rydberg states.

A more detailed description of our model can be found in Ref.\,\cite{sassmannshausen2015}.

\subsection{Pulsed single-colour two-photon excitation}
\label{sec:pulsed}
\subsubsection{Experimental}
The measurements were carried out under similar conditions as decribed in Section\,\ref{experimental}. The continuous-wave output of the ring dye laser is pulse amplified in two dye cells pumped by the second harmonic of a pulsed Nd:YAG laser. The pulse-amplified radiation is frequency doubled in a BBO crystal, and UV pulses of up to 100\,$\upmu$J pulse energy, a pulse duration of 4.4\,ns and a Fourier-transform-limited frequency bandwidth of 130\,MHz are obtained. The UV-laser beam is focused to a beam waist radius of $\sim150$\,$\upmu$m, resulting in intensities of up to 10$^8$\,W/cm$^2$. For these experiments we use samples of ultracold Cs atoms with a density of up to 10$^{12}$\,cm$^{-3}$ and a temperature of $40 \mu$K as released from a far-detuned crossed optical dipole trap (1064 nm, 10 W, 80\,$\upmu$m beam waist radius). Prior to the application of the UV-laser pulse, the atoms in the optical dipole trap are selectively pumped to the 6s$_{1/2}, F=4$ hyperfine level with a focused repumping laser. The power of the trapping laser is reduced to its minimum of 300\,mW prior to the application of the UV-laser pulse used to excite the atoms to $n$p$_{3/2}$ Rydberg states. At this low power, the trapping laser radiation has a negligible influence on the Rydberg atoms, for which we estimate a ponderomotive shift of about 150\,kHz only. A delay of 5\,$\upmu$s is introduced between the UV-laser pulse and the pulsed electric field used for field ionisation and ion extraction.
\begin{figure}[t!]
\center
\resizebox{0.32\linewidth}{!}{%
 \includegraphics{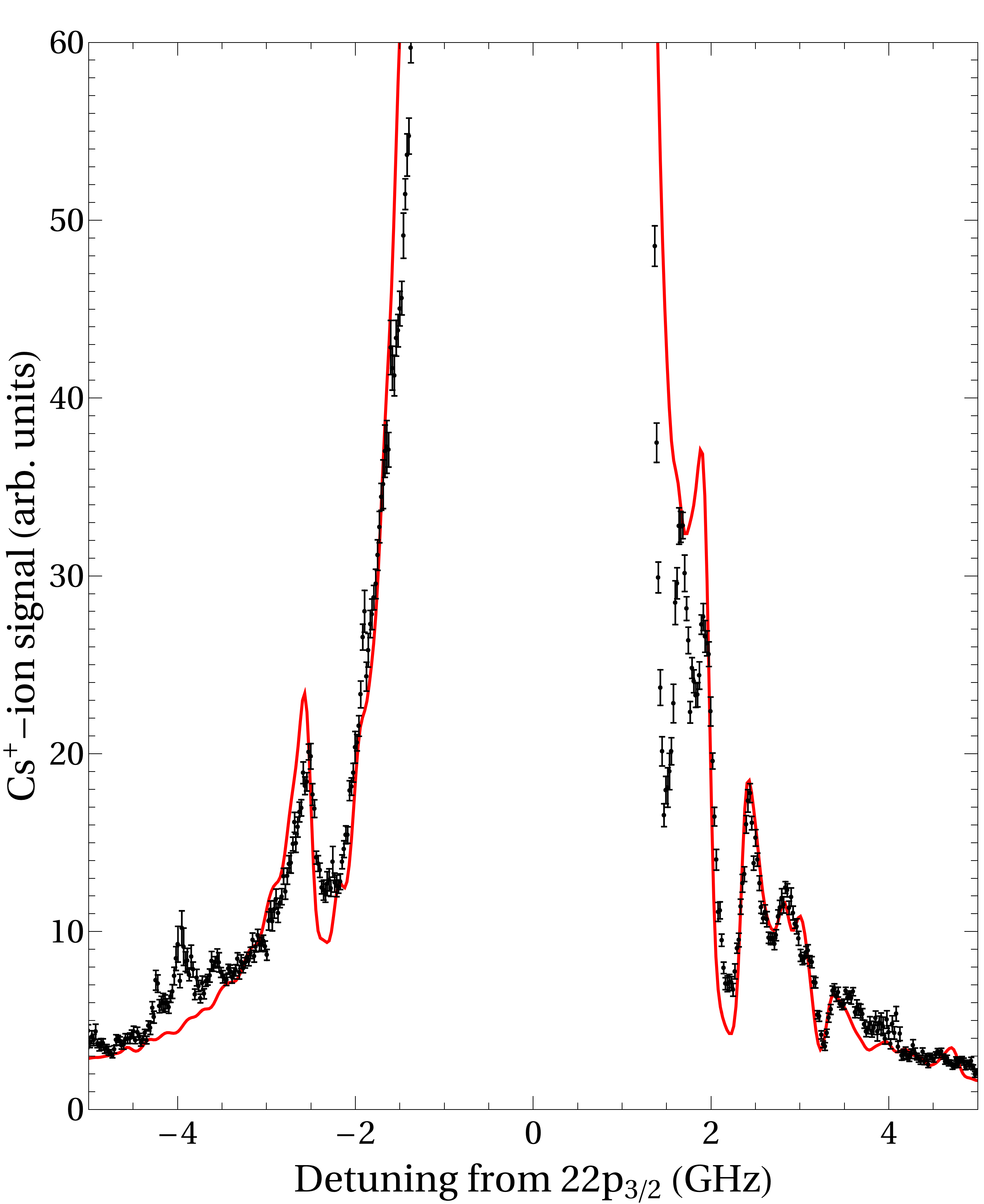}}
\resizebox{0.32\linewidth}{!}{%
 \includegraphics{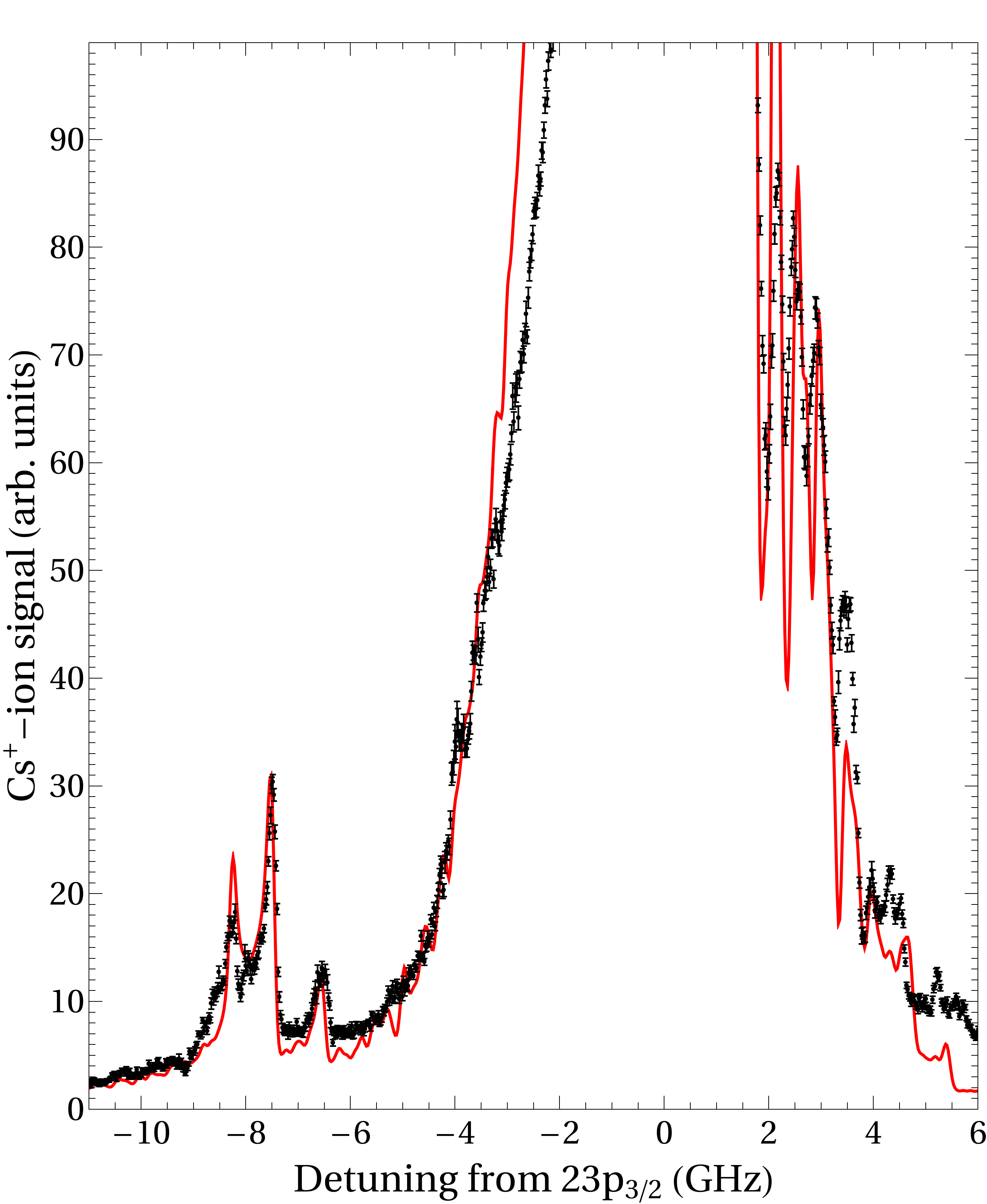}}
\resizebox{0.32\linewidth}{!}{%
 \includegraphics{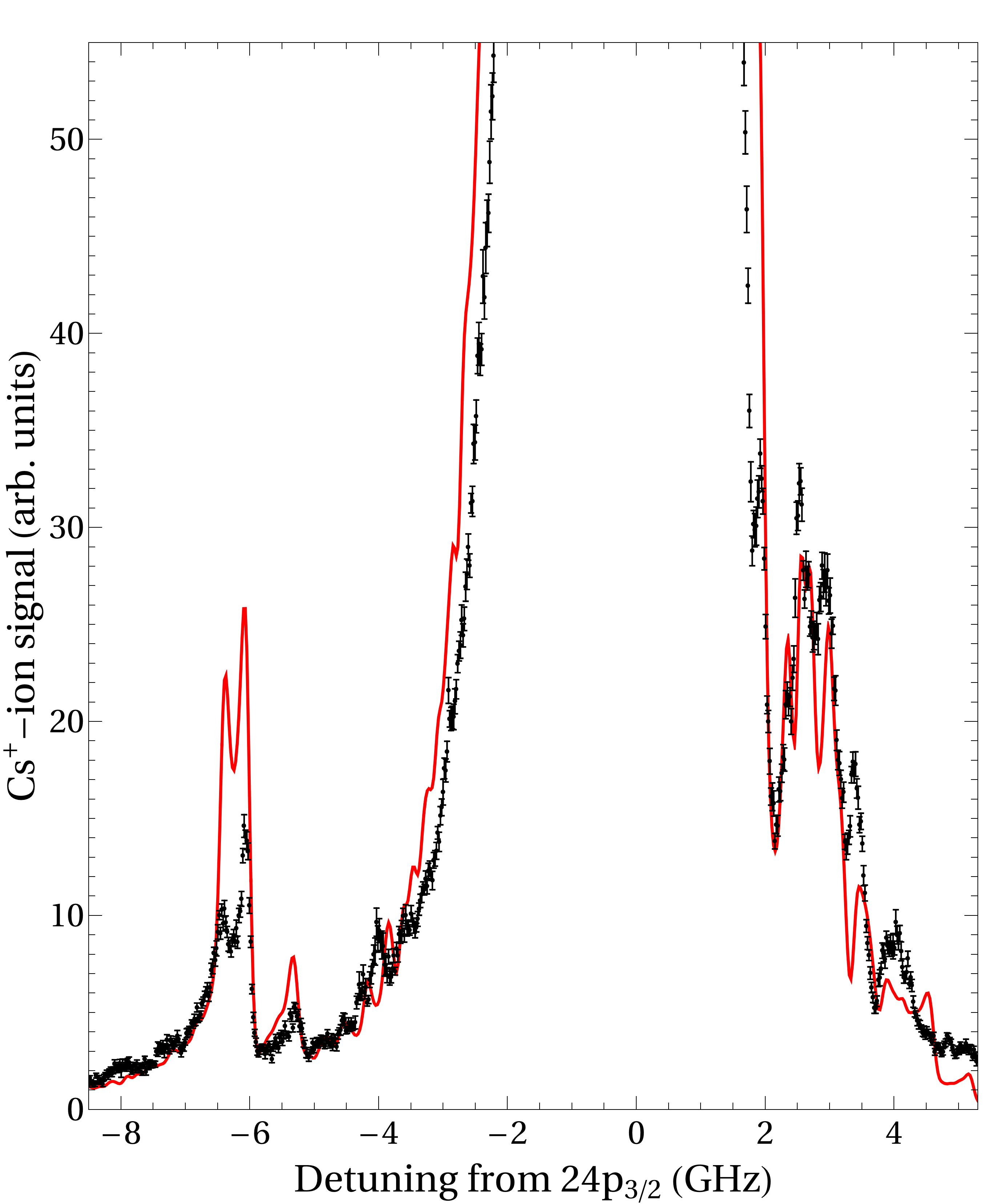}}
\resizebox{0.32\linewidth}{!}{%
 \includegraphics{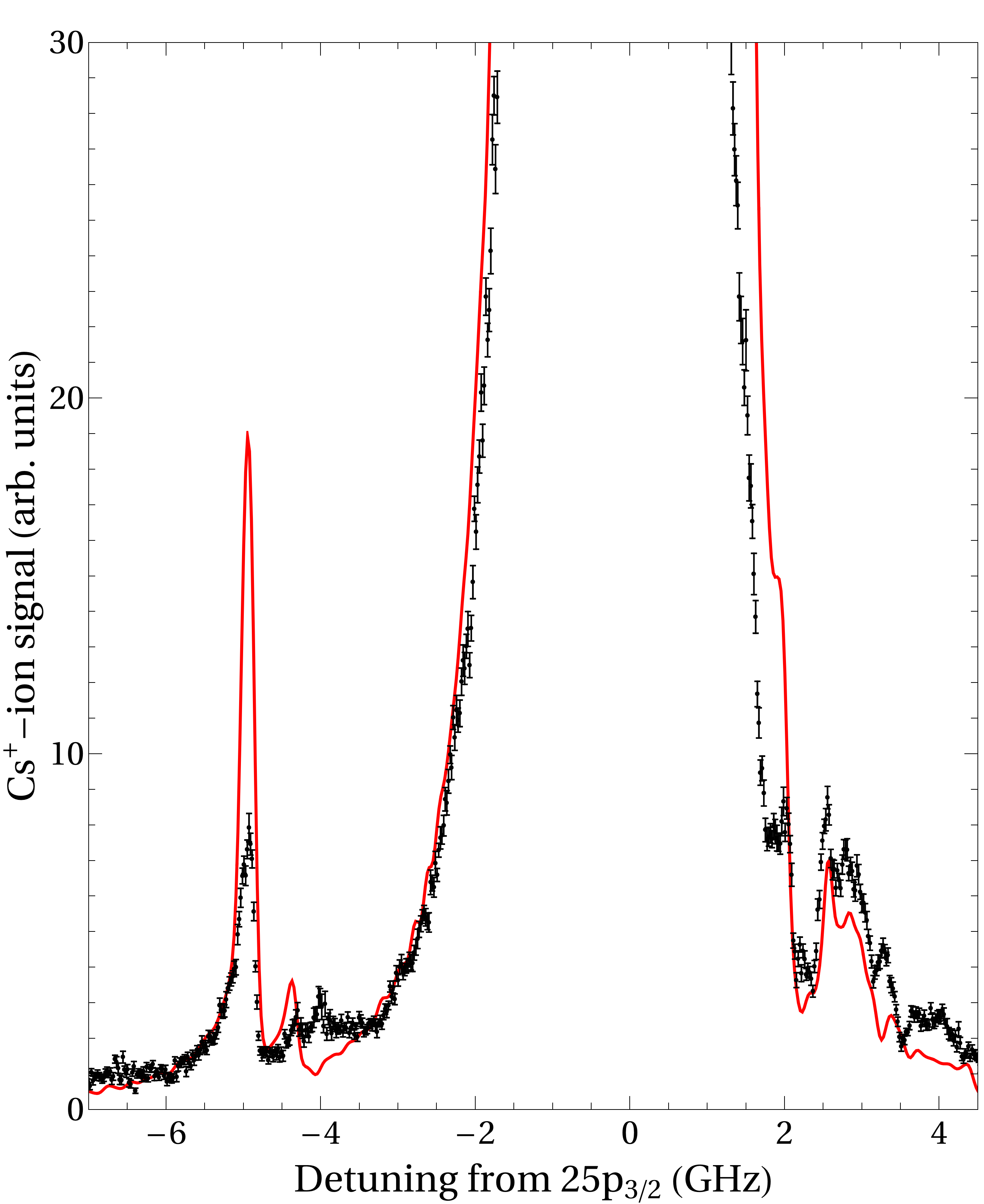}}
\resizebox{0.32\linewidth}{!}{%
 \includegraphics{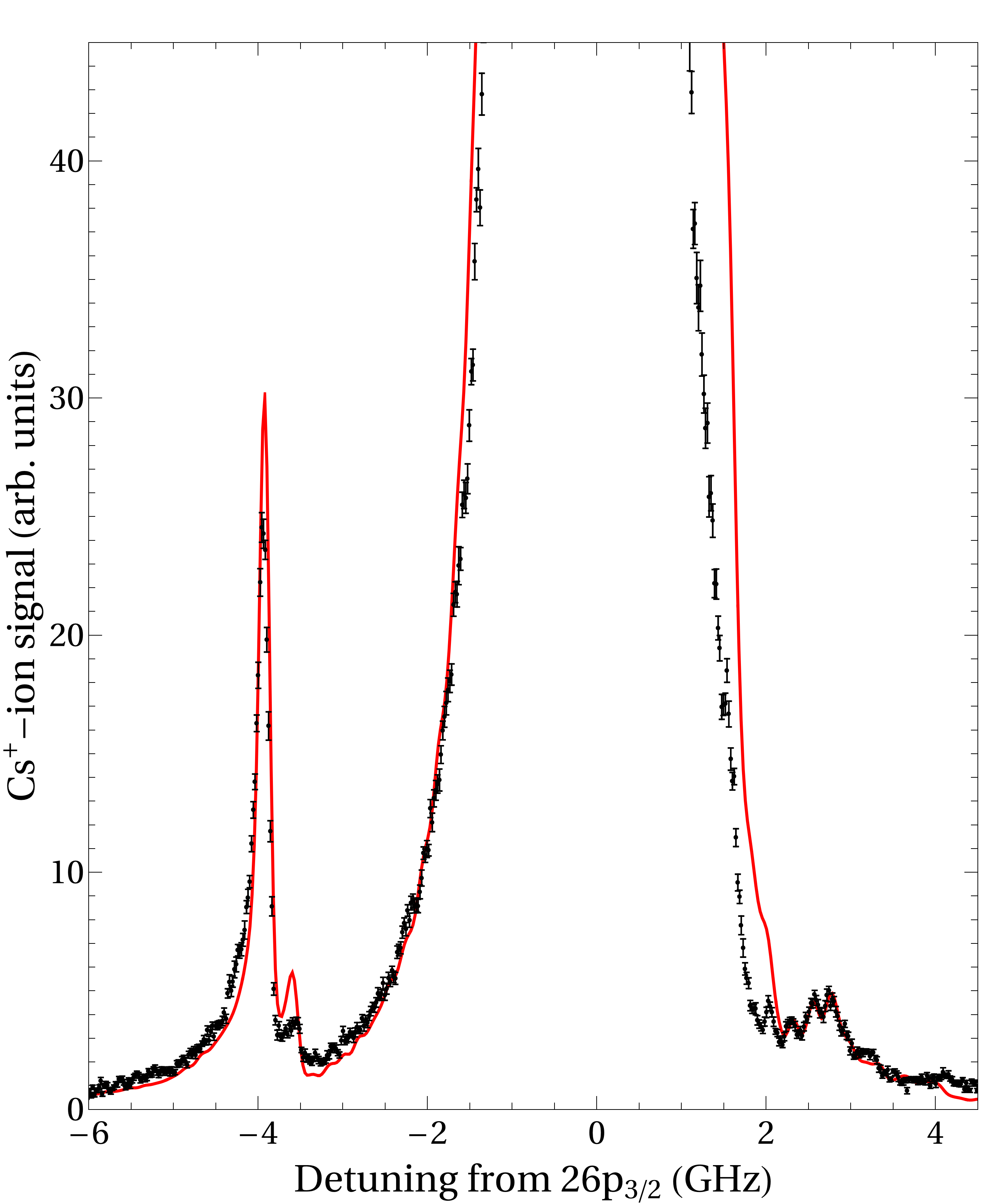}}
\resizebox{0.32\linewidth}{!}{%
 \includegraphics{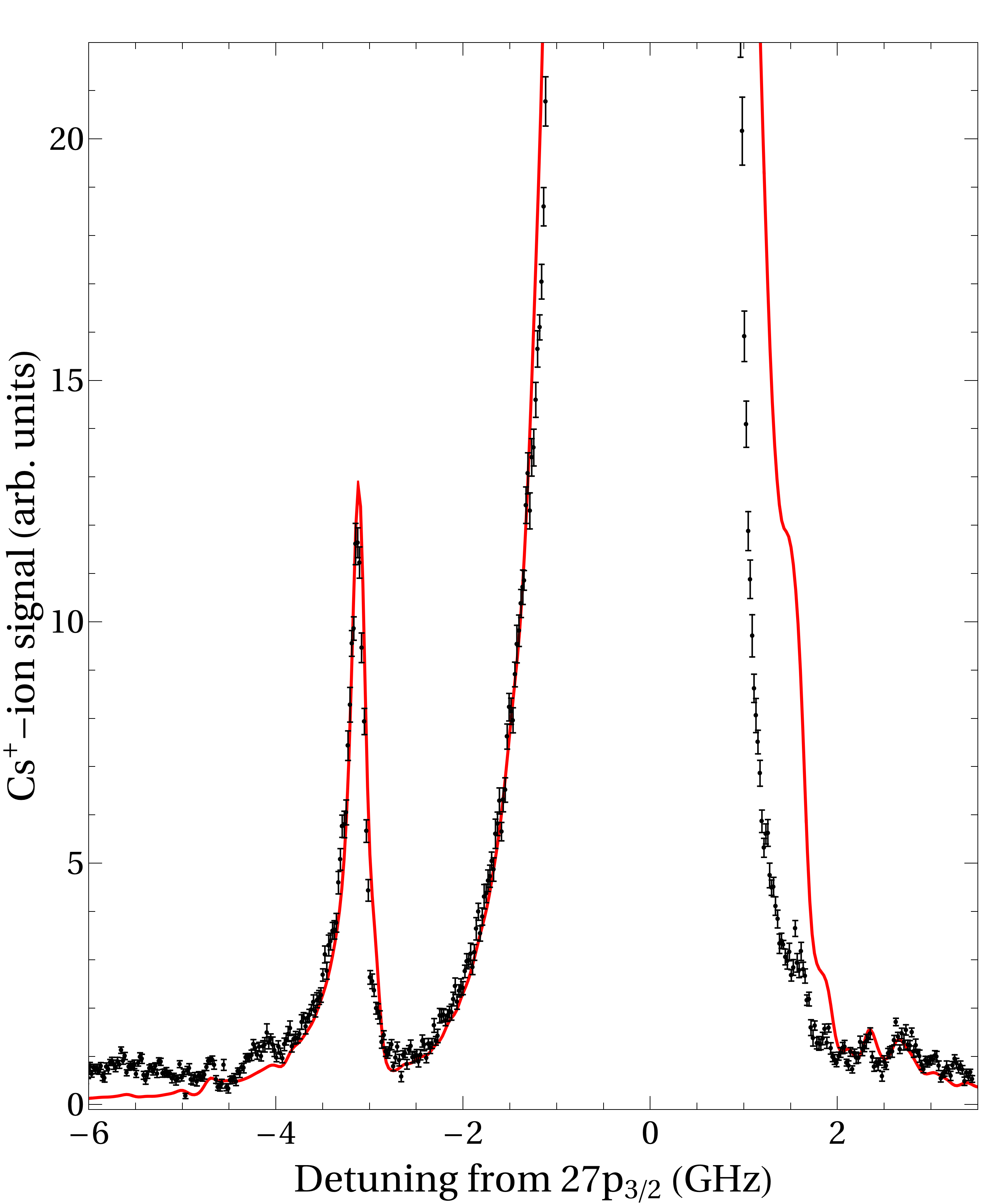}}
\caption{Pulsed UV-laser excitation spectra of interacting Rydberg-atom-pair states in the vicinity of $n$p$_{3/2} \leftarrow 6$s$_{1/2}$ transitions in Cs for $n=22-27$. The Cs$^+$-ion signal produced spontaneously during the 5\,$\upmu$s between the UV pulse and electric extraction pulse is shown as black dots with error bars representing the standard deviation of the average of many experimental cycles. Spectra simulated on the basis of the long-range interaction model taking into account terms up to the octupole-octupole contribution are shown as solid red lines (see text for details).}
\label{fig:6}
\end{figure}

\subsubsection{Results}
Typical spectra close to $n$p$_{3/2} \leftarrow 6$s$_{1/2}$ transitions for $21 < n < 28$ recorded under these conditions are presented in Fig.\,\ref{fig:6}. The black dots represent the measured spontaneous Cs$^+$-ion signals and the red traces represent the simulated spectra. The simulations do not rely on any adjustable parameter (apart from an overall intensity-scaling factor) and accurately reproduce all sharp spectral structures of the experimental data. The resonances on the low-frequency side of  $n = 23 - 27$ $n$p$_{3/2} \leftarrow 6$s$_{1/2}$ transitions correspond to the excitation of Rydberg-atom-pair states correlated to $n$s$_{1/2}(n+1)$s$_{1/2}$ dissociation asymptotes\,\cite{sassmannshausen2015}. The dominant interaction giving rise to these resonances is the dipole-dipole interaction. Two exemplary potential curves are depicted in\,Fig.\,\ref{fig:5}b and c for spectral regions around the 23p$_{3/2}$23p$_{3/2}$ and the 26p$_{3/2}$26p$_{3/2}$ pair-dissociation asymptote, respectively. The resonance on the low-frequency side of the 22p$_{3/2} \leftarrow 6$s$_{1/2}$ transition originates from the 22s$_{1/2}$19f$_{J}$ asymptote located at a detuning of $\sim2.5$\,GHz below the 22p$_{3/2}$22p$_{3/2}$ asymptote (see Fig.\,\ref{fig:5}c and Fig.\,\ref{fig:6}). The observation of this resonance is the result of the dipole-quadrupole interaction\,\cite{deiglmayr2014}. The dipole-quadrupole interaction does not conserve the electronic parity of the Rydberg-atom-pair system. Consequently, its observation betrays the coupling of electronic and nuclear degrees of freedom. This coupling is facilitated in long-range Rydberg macrodimers because their rotational levels are quasi-degenerate, \emph{i.e.}, the spacings between the low-lying molecular rotational levels are smaller than their natural linewidths. This situation is very similar to the long-range Rydberg molecules bound by low-energy electron-Cs scattering described in Section\,2.3.

The rich spectral features observed on the high-frequency side of the $n$p$_{3/2} \leftarrow 6$s$_{1/2}$ atomic resonances can only be accurately simulated after inclusion of high-$l$ states into the pair-state basis set and of terms up to the octupole-octupole-interaction operator in the long-range Hamiltionian. The simulations reveal that these features result from avoided crossings between repulsive $n$p$_{3/2}n$p$_{3/2}$ potential curves and attractive curves that correlate to higher-lying $(n-2)$d$_{3/2}n$d$_{3/2}$ and $(n-4)$f$_J(n-3)$f$_J$ pair-dissociation asymptotes. These avoided crossings on the high-frequency side of 23p$_{3/2}$23p$_{3/2}$ and 26p$_{3/2}$26p$_{3/2}$ asymptotes are depicted in Fig.\,\ref{fig:5}b and c and occur at internuclear distances of only about 200-400\,nm. Because there are no molecular potential curves with significant $n$p$_{3/2}n$p$_{3/2}$ character at the positions of the crossings, they can be directly linked to intensity minima in the experimental and simulated spectra.

We can observe these resonances using pulsed field ionisation, although a higher sensitivity and a better signal-to-noise ratio are obtained when monitoring atomic ions produced by spontaneous ionisation of the sample during the delay between the laser pulse and the electric-field pulse. Possible mechanisms for the decay of Rydberg-atom-pair states are discussed in Section\,\ref{sec:decay}.

\subsection{Resonant two-colour two-photon excitation}
\label{sec:seed}
The sequential resonant excitation we use to access interacting Rydberg-atom-pair states
\begin{equation}
\label{eq:scheme}
nl_Jnl_J \xleftarrow{\nu'=\nu-\mathrm{\Delta} E(R)/h} nl_J6\mathrm{s}_{1/2} \xleftarrow{\nu} 6\mathrm{s}_{1/2}6\mathrm{s}_{1/2}
\end{equation}
requires radiation of two different frequencies $\nu$ and $\nu'$ because the Rydberg-atom-pair states are shifted from the positions of the pair-dissociation asymptotes. The first experiments implementing this type of excitation scheme were performed by Reinhard \textit{et al.}\,\cite{Reinhard2008} with the goal of characterising the Rydberg-excitation-blockade effect. A broadening of the Rydberg transitions was observed depending on the first resonant laser pulse. The Rydberg excitation by the second, detuned laser was found to be facilitated by the first, resonant laser because atoms located at the distance $R_{\rm fac}$\,\cite{lesanovsky2014} from Rydberg atoms produced by the first laser experience a shift of their transition frequency into resonance with the frequency of the second laser. This anti-blockade effect\,\cite{ates2007,amthor2010} can lead to non-Poissonian counting statistics and bimodal counting distributions in off-resonant Rydberg excitation\,\cite{schempp2014,malossi2014} and to the formation of spatially correlated Rydberg aggregates\,\cite{hu2013,lesanovsky2014}.

Here, we present experiments on Cs Rydberg atoms using a sequential two-colour excitation where one photon is resonant with an intermediate singly-excited $n$p$_{3/2}$6s$_{1/2}$ state (see Fig.\,\ref{fig:1}c and Eq.\,\ref{eq:scheme}). Our main motivation is to study interacting Rydberg-atom-pair states at higher resolution than is achievable in one-colour two-photon excitation schemes. The results presented in this section demonstrate that a narrow-bandwidth cw laser can be used to drive the second transition. In this way, the spectral resolution can be improved by almost two orders of magnitude and Rydberg-atom-pair states can be excited at well-defined internuclear separations.

\subsubsection{Experimental}
The first transition (see Eq.\,\ref{eq:scheme} and Fig.\,\ref{fig:1}c) is driven by a pulse-amplified and frequency-doubled diode laser (Toptica DL-100 pro). The output of the diode laser is sent through two dye cells pumped by the second harmonic of a seeded Nd:YAG laser for amplification, and through a BBO crystal for second-harmonic generation. In this way, we obtain UV pulses of 4.4\,ns duration with a Fourier-transform-limited bandwidth of $\sim130$\,MHz, very similar to the UV-laser pulses used for the experiments described in Section\,\ref{sec:pulsed}. The frequency of this ``short-pulse'' UV laser is kept resonant with a selected atomic $n$p$_{3/2} \leftarrow 6$s$_{1/2}$ transition and its power is reduced to levels corresponding to the excitation of typically $\sim40$ atoms, called ``seed'' Rydberg atoms in the following, per pulse. The second laser is a single-mode continuous-wave frequency-doubled ring dye laser, the characteristics of which are described in section\,\ref{sec:trilos}. We switch the light from this laser with an acousto-optical modulator and generate long laser pulses (typically 15\,$\upmu$s duration) which interact with the ultracold sample containing the ``seed'' Rydberg atoms. Both lasers are focused into the ultracold cloud of Cs atoms (beam waist radii of $\sim150$\,$\upmu$m), where the two laser beams cross at right angles. We extract spontaneously produced ions and field ionise remaining Rydberg atoms with a pulsed electric field delayed with respect to the second laser pulse. The Cs$^+$ ions produced by spontaneous ionisation or pulsed field ionisation are detected mass-selectively with an MCP detector at the end of a time-of-flight tube. We present here spectra obtained by monitoring the Cs$^+$-ion yield as a function of the frequency of the long-pulse UV laser.

\subsubsection{Results}
An example of a spectrum recorded with the resonant two-colour excitation scheme described above is presented in Fig.\,\ref{fig:1}c. The spectral resolution is $\sim6$\,MHz, about twenty times better than in the spectrum recorded with the short-pulse UV laser (see Fig.\,\ref{fig:1}a). The experimental spectrum is accurately reproduced over a broad spectral range around the 32s$_{1/2}33$s$_{1/2}$ Rydberg-atom-pair resonance by a spectrum simulated using a version of Eq.\,\ref{eq:lineprofile} adapted to account for the resonant one-photon excitation of the pair state from the intermediate 32p$_{3/2}$6s$_{1/2}$ level. However, the intensity of the simulated spectrum is systematically too weak at small detunings (within $\sim1$\,GHz) of the long-pulse UV-laser frequency from the atomic 32p$_{3/2} \leftarrow 6$s$_{1/2}$ resonance. The observation of much larger signals in the experiment at frequencies just below the atomic transition frequency indicates that the second laser pulse excites more than one atom per seed atom. This form of ``aggregation'' of Rydberg-atom excitation has been referred to as ``avalanche'' excitation in previous work\,\cite{simonelli2016} and is a characteristic feature of the resonant two-colour excitation discussed in this section.
\begin{figure}[t!]
\center
\resizebox{0.72\linewidth}{!}{%
 \includegraphics{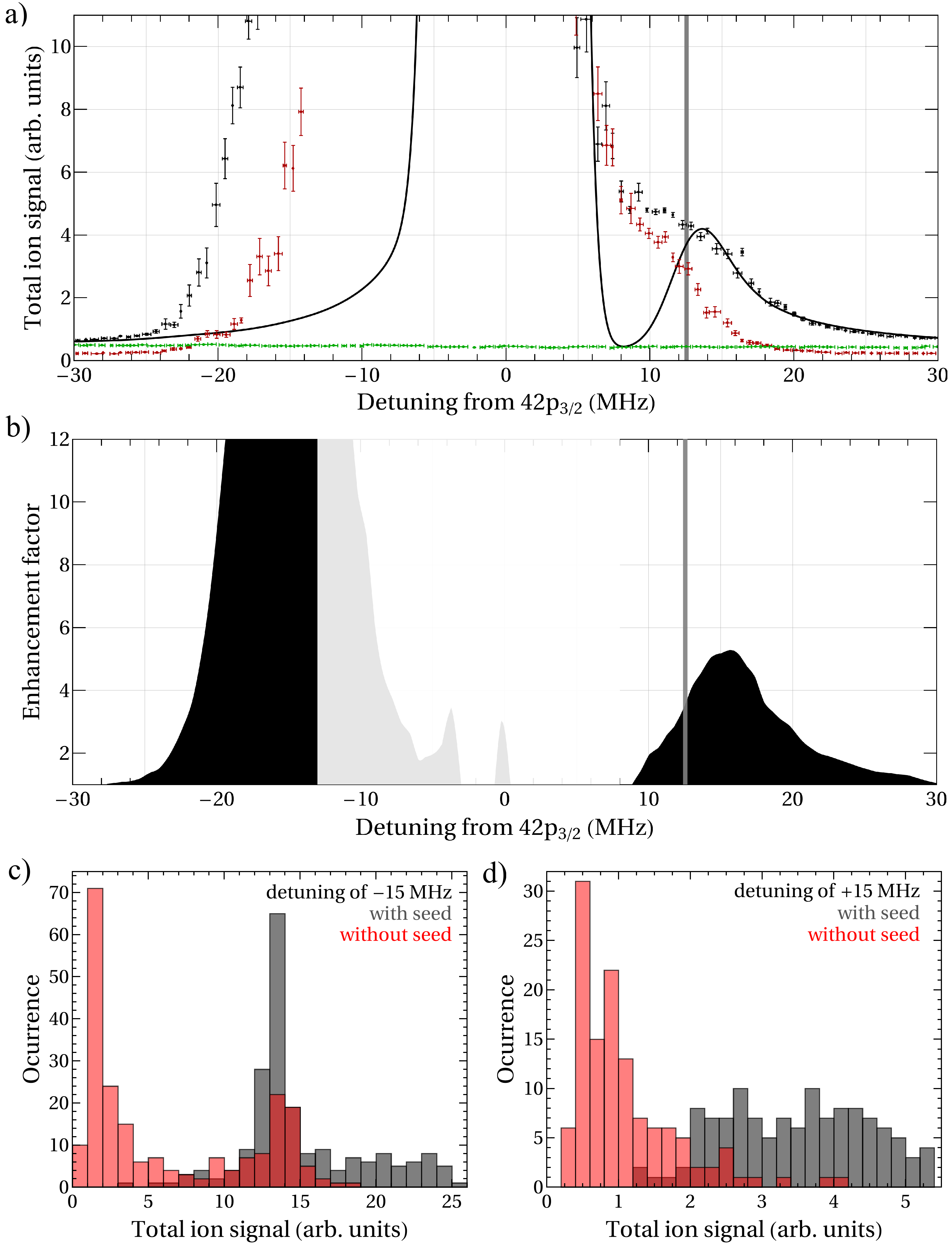}}
\caption{(a) Two-colour UV-laser spectrum recorded in the vicinity of the atomic 42p$_{3/2} \leftarrow 6$s$_{1/2}$ transition (black dots). The green dots represent the background signal caused by the short-pulse UV laser alone with its frequency fixed at the position of the 42p$_{3/2} \leftarrow 6$s$_{1/2}$ transition. The spectrum obtained without any seed Rydberg atoms, \emph{i.e.} without the short-pulse laser, is presented as dark red dots. The vertical grey bar indicates the asymptotic energy of the 42s$_{1/2}43$s$_{1/2}$ pair state. The full line represents a spectrum calculated with the model of the multipole-multipole interactions between Rydberg atoms (see section\,3.1). (b) Difference between the two-colour spectrum and the long-pulse UV-laser spectrum presented as black and red dots in (a), respectively, divided by the seed Rydberg signal (green dots in (a)) as a function of the laser detuning from the 42p$_{3/2} \leftarrow 6$s$_{1/2}$ transition (black). The counting distributions at detunings of -15\,MHz and +15\,MHz for the excitation without (red bars) and with (black bars) seed Rydberg atoms are presented in panels (c) and (d), respectively.}
\label{fig:7}
\end{figure}

\begin{figure}[t!]
\center
\resizebox{0.8\linewidth}{!}{%
 \includegraphics{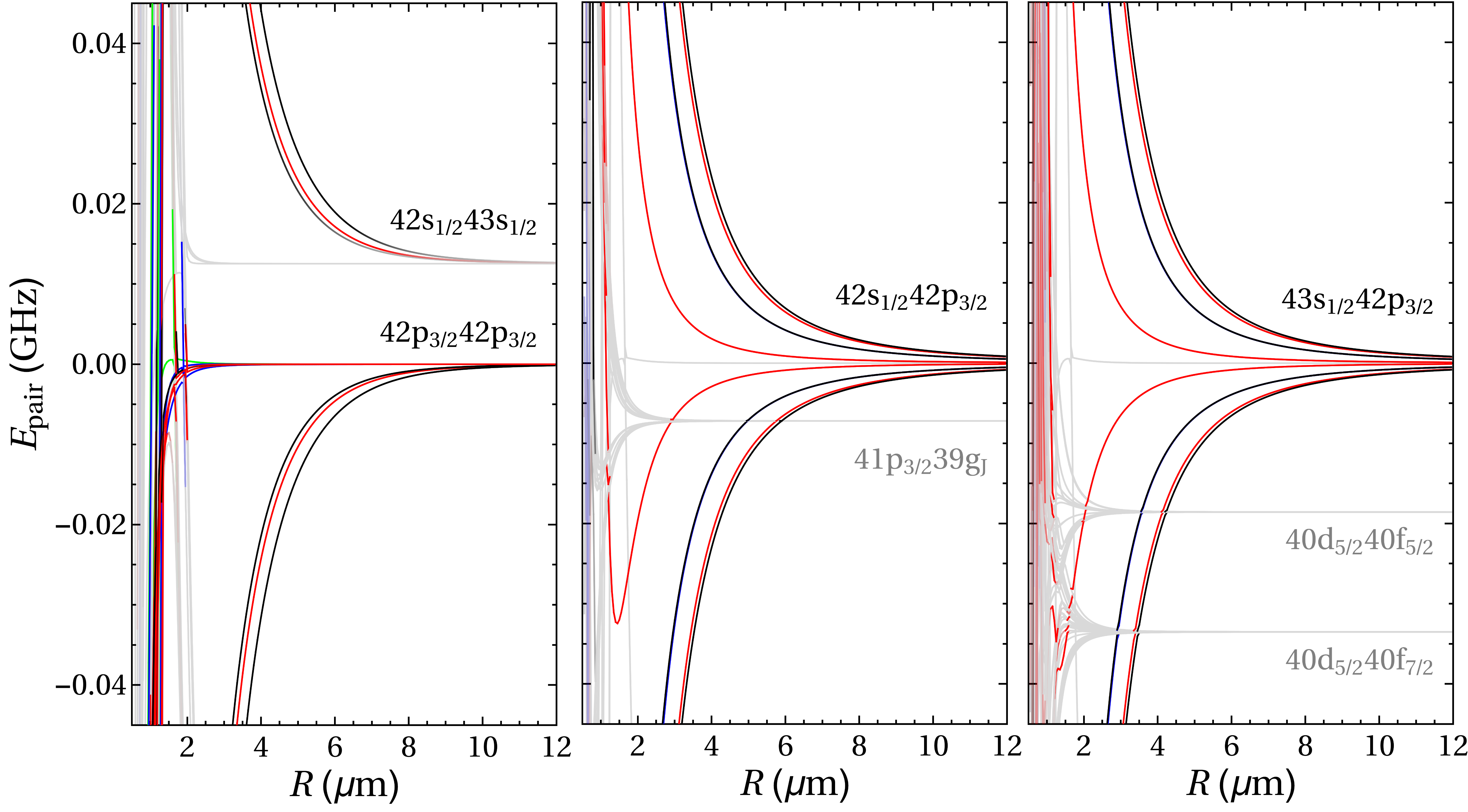}}
\caption{Long-range potential-energy curves in an energy rage of $\pm 40 \cdot h$\,MHz around the 42p$_{3/2}$42p$_{3/2}$ (left-hand side), the 42s$_{1/2}$42p$_{3/2}$ (middle) and the 43s$_{1/2}$42p$_{3/2}$ (right-hand side) Rydberg-pair-dissociation asymptotes. The colour shading of the molecular potential curves represent the admixture of the asymptotic level, which is placed at the origin of the energy scale. Full colour denotes 1\,\% character of the specific asymptotic state, and black and red traces indicate molecular states with $\Omega=0,1$, respectively.}
\label{fig:8}
\end{figure}

To study the effect of the off-resonant sequential ``avalanche'' Rydberg excitation and the formation of Rydberg-atom aggregates in more detail, we have concentrated on the region near the atomic 42p$_{3/2} \leftarrow 6$s$_{1/2}$ transition in Cs (see Fig.\,\ref{fig:7}). At this value of $n$, the energy spacing between $n$p$_{3/2}n$p$_{3/2}$ and $n$s$_{1/2}(n+1)$s$_{1/2}$ Rydberg-atom-pair states is only $12.5 \cdot h$\,MHz, leading to significant state-mixing and level shifts already at large internuclear separations\,\cite{viteau2008}. The relevant potential curves are depicted in Fig.\,\ref{fig:8}. In the experiment, the frequency of the short-pulse UV laser was tuned to the position of the 42p$_{3/2} \leftarrow 6$s$_{1/2}$ transition. Consequently, the potential curves needed to describe the excitation with the long-pulse UV laser are correlated to the Cs\,42p$_{3/2}$ + Cs\,42p$_{3/2}$ and the Cs\,42s$_{1/2}$ + Cs\,43s$_{1/2}$ dissociation asymptotes (see left-hand-side panel of Fig.\,\ref{fig:8}). The molecular potential curves that correlate with the 42s$_{1/2}$43s$_{1/2}$ functions are repulsive and the dipole-dipole 42p$_{3/2}$42p$_{3/2}-$42s$_{1/2}$43s$_{1/2}$ interaction implies the occurence of a weak resonance with a blue-degraded line shape close to the position of the 42s$_{1/2}$43s$_{1/2}$ asymptote on the high-frequency side of the atomic 42p$_{3/2} \leftarrow 6$s$_{1/2}$ transition. The dipole-dipole interaction also implies that the molecular functions correlated to the 42p$_{3/2}$42p$_{3/2}$ Rydberg-atom-pair state are attractive and lead to a strong red-degraded molecular resonance on the low-frequency side of the 42p$_{3/2} \leftarrow 6$s$_{1/2}$ atomic resonance.

The experimental spectra are presented in Fig.\,\ref{fig:7}a. The frequency-independent signal presented in green was measured with the short-pulse UV laser alone and corresponds to the pulsed field ionisation of the ``seed'' 42p$_{3/2}$ Rydberg atoms. The red trace was measured with the long-pulse UV laser alone, and the black trace was recorded with both lasers. As explained in the experimental section, the lasers did not overlap in time, and the second laser pulse was applied with a small delay of $\sim5$\,ns after the first laser pulse. In the spectrum shown in black, a partially resolved resonance is observed on the high-frequency side of the atomic resonance, at the position of the 42s$_{1/2}$43s$_{1/2}$ asymptote (see vertical grey line in Fig.\,\ref{fig:7}). This resonance is also observed in the simulated spectrum (solid black trace in Fig.\,\ref{fig:7}), that was obtained with the potential functions depicted in the left-hand-side panel of Fig.\,\ref{fig:8}. In the experimental trace, the resonance is not as well resolved as in the simulation because of partial overlap with the broadened atomic transition.

Whenever the experimental black trace in Fig.\,\ref{fig:7} is larger than the sum of the spectra shown in green and red, the additional signal can unambiguously be attributed to the two-colour two-photon excitation of Rydberg-atom-pair states. The additional Rydberg excitations are visualised in Fig.\,\ref{fig:7}b, which depicts the difference between the spectrum measured with and without seed Rydberg atoms normalised to the seed-Rydberg-atom signal. This spectrum represents the enhancement factor of the photoexcitation probability induced by the seed Rydberg atoms. When this factor becomes greater than two, more than one additional Rydberg atom is excited for each seed Rydberg atom, which is indicative of an aggregation of Rydberg excitations. At the 42s$_{1/2}$43s$_{1/2}$ pair resonance on the high-frequency side of the atomic transition, enhancement factors up to 4 are observed. The excitation of the 42s$_{1/2}$43s$_{1/2}$ pair resonance therefore facilitates additional Rydberg excitations during the 15-$\upmu$s-long UV-laser pulse. An exact theoretical description would require the calculation of few-body--Rydberg-atom states, which is beyond the scope of this article. The relevant potentials for the excitation of a third Rydberg atom close to a Rydberg-atom-pair state will, however, resemble those depicted for 42s$_{1/2}$42p$_{3/2}$ and 43s$_{1/2}$42p$_{3/2}$ Rydberg-atom-pair states states in Fig.\,\ref{fig:8}. The always-resonant $n$s$_{1/2}n'$p$_{3/2}\leftrightarrow n'$p$_{3/2}n$s$_{1/2}$ dipole-dipole interaction in these dimers gives rise to long-range interactions (detunings of 15\,MHz correspond to internuclear distances of $\sim5$\,$\upmu$m), and leads to the formation of small Rydberg aggregates after the initial pair-state excitation. The state mixing induced by this interaction will also strongly modify the interactions in the 42s$_{1/2}$43s$_{1/2}$ pair state. Consequently, on the high-frequency side of the atomic transition, the system should be described as a three- or few-body state with the p character fully ``delocalised'' over all interacting Rydberg atoms. The experimental data indicates that these interactions lead to the formation of small Rydberg aggregates, with enhancement factors much smaller than observed on the low-frequency side of the atomic 42p$_{3/2} \leftarrow 6$s$_{1/2}$ transition (see Fig.\,\ref{fig:7}b).

On the low-frequency side, the pair states carry mainly 42p$_{3/2}$42p$_{3/2}$ character. The very large enhancement factors of much more than 10 for detunings between $-10$ and $-15$\,MHz correspond to an excitation avalanche. Each 42p$_{3/2}$ Rydberg atom excited with the long-pulse UV laser also acts as an additional seed atom facilitating the excitation of further neighbouring atoms, which results in a spatially correlated many-body state\,\cite{lesanovsky2014}. This process is also observed with the long-pulse laser alone as revealed already by the underestimation of the signals at detunings between $-20$ and $-10$\,MHz in the simulated spectra (compare red dots and solid black line in Fig.\,\ref{fig:7}a). The counting distributions recorded at detunings of $-15$ and $+15$\,MHz also reveal this effect of the excitation avalanche. The counting distribution measured without seed atoms is shown in red in Fig.\,\ref{fig:7}c and d and is bimodal for the Rydberg excitation at a detuning of $-15$\,MHz. In this case, the majority of the experimental cycles leads to the formation of only a few Rydberg-atom pairs, but a few cycles give rise to an ionisation signal that is more than 10 times stronger, and of the same strength as observed when the short-pulse UV laser is turned on. The conditions for an excitation avalanche are thus only occasionally realised. It thus appears that not all seed atoms are effective in inducing an excitation avalanche, but those that are effective lead to very similar enhancement of the excitation. The behaviour is reminiscent of that of a saturable autocatalytic process, and suggest that the process terminates when the entire sample has been excited. Assuming that the initial excitation occurs at the centre of the 30-$\upmu$m-diametre cloud of ultracold Cs atoms and taking a facilitation radius of 5\,$\upmu$m for a detuning of $-15$\,MHz (derived from the potential functions depicted in Fig.\,\ref{fig:8}), the estimated enhancement factor corresponding to the avalanche excitation of the entire sample is $\sim100$.
At a positive detuning of $+15$\,MHz, the counting distributions do not reveal any sign of an excitation avalanche, even though the enhancement factor of $\sim4$ indicates the formation of small Rydberg-atom aggregates. Additional experimental and theoretical work is needed to fully understand these intriguing observations.

The main advantage of the successive two-colour excitation sequence for the study of interacting Rydberg-atom-pair states is the increased frequency resolution demonstrated in this section (see also Fig.\,\ref{fig:1}). Another advantage, not exploited here, is the possibility of accessing pair states consisting of Rydberg atoms in different angular-momentum states (such as $n$s$n$p pair states), \emph{e.g.} by preparing the seed Rydberg atoms using a two-photon transition. Finally, a two-colour excitation scheme is well suited to study the formation of larger Rydberg aggregates under conditions where the excitation of many atoms is facilitated by a single or only very few seed Rydberg atoms, as demonstrated above.

\section{Decay channels of molecular states}
\label{sec:decay}
\begin{figure}[t!]
\center
\resizebox{\linewidth}{!}{%
 \includegraphics{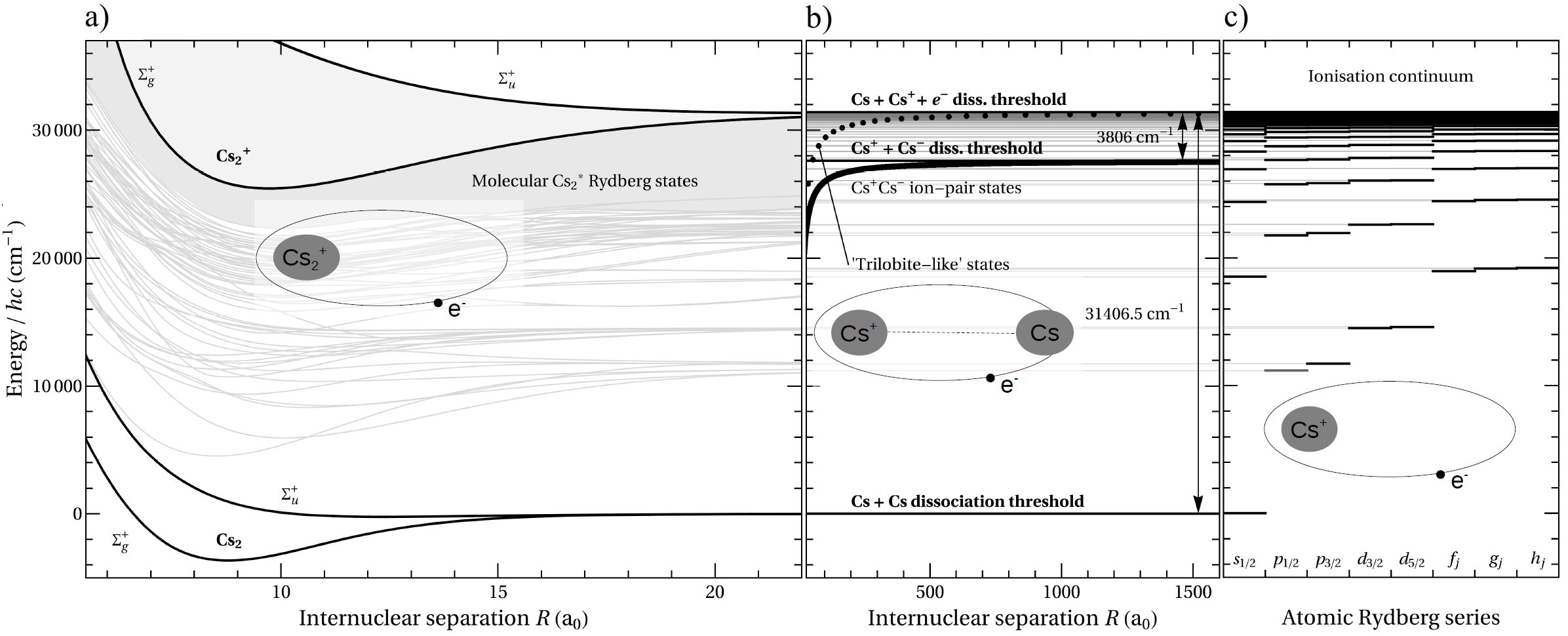}}
\caption{Overview of molecular potential functions of Cs$_2$ and Cs$_2^+$ below the first dissociative-ionisation threshold of Cs$_2$ (Cs$^+~{^1}{\rm S}_0 + {\rm Cs}~6{\rm s}_{1/2}$) in different ranges of internuclear separation. The electronic structure of isolated Cs atoms consists of Rydberg series of increasing orbital angular momentum and is depicted in panel (c). Panel (a) shows molecular potentials of covalently bound Cs$_2$ molecules. The lowest two ground-state potentials of $g$ and $u$ symmetry of Cs$_2$ and Cs$_2^+$ are depicted in black, whereas the excited states of the Cs$_2$ dimer having Rydberg character are shown in grey (calculated potentials taken from Ref.\,\cite{spiess1989}). At high $n$ values, the state density is high and the level structure is represented as a pseudo-continuum (grey background). Panel (b) focuses on the long-range regime where orbital overlap of low-lying states of the two atoms is negligible and the Cs$_2$ ground-state potential is essentially flat. The thick black line corresponds to the Coulomb potential of the Cs$^+$Cs$^-$ ion pair and the black dots give the energy and average internuclear separation of the long-range Rydberg molecules discussed in section 2. All energies are given with respect to the ${\rm Cs}$\,$6{\rm s}_{1/2} + {\rm Cs}$\,$6{\rm s}_{1/2}$ dissociation threshold.}
\label{fig:9}
\end{figure}
\begin{figure}[t!]
\center
\resizebox{0.85\linewidth}{!}{%
 \includegraphics{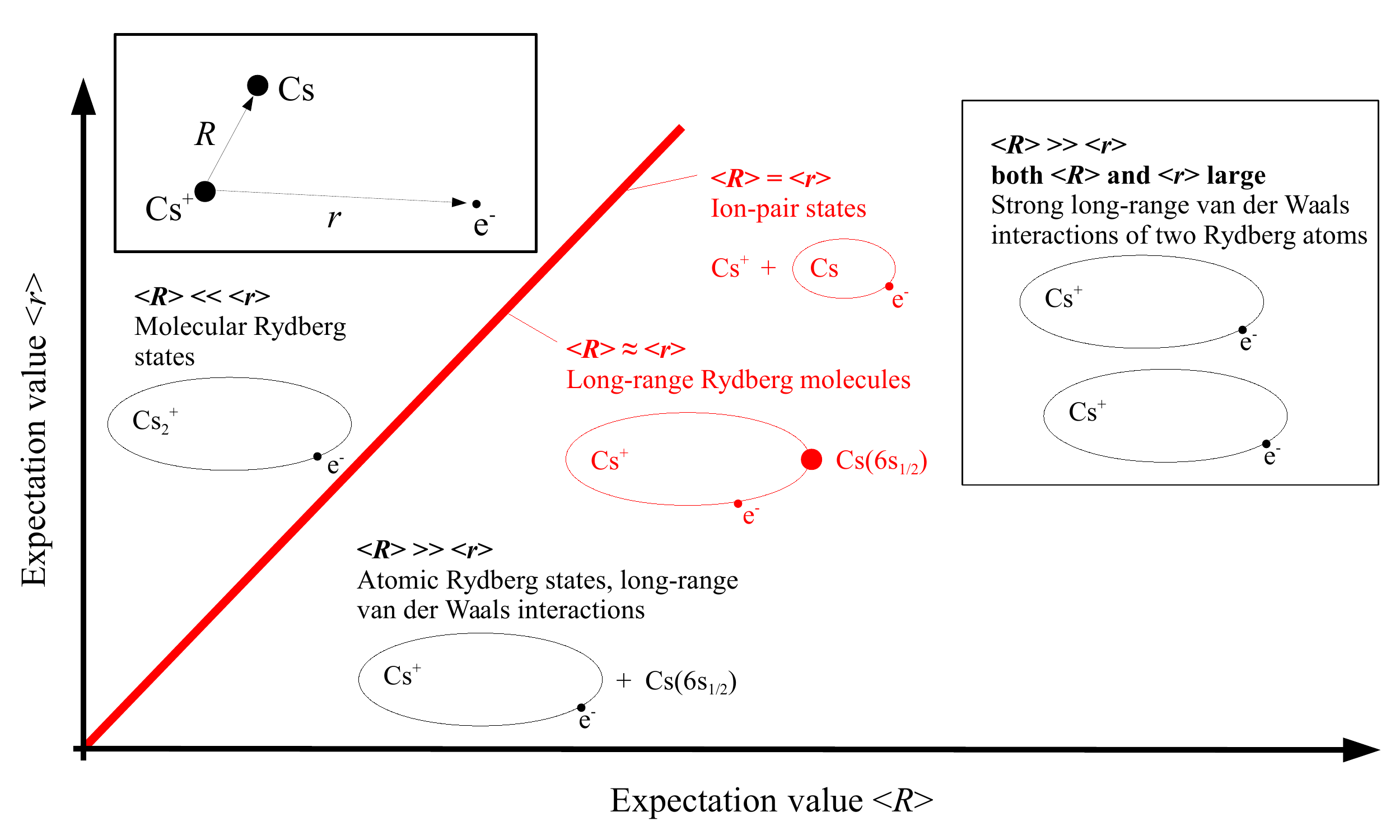}}
\caption{Schematic illustration of different molecular regimes as a function of the expectation value of the internuclear distance, $\left<R\right>$, and the expectation value of the distance of one electron to the Cs$^+$ ion, $\left<r\right>$.}
\label{fig:10}
\end{figure}

The two types of long-range molecular states discussed in this article are detected through their ionic decay products. The long-range Rydberg molecules were found to decay into molecular Cs$_2^+$ ions, whereas the interacting Rydberg-atom-pair states decayed into atomic Cs$^+$ ions. To evaluate which decay channels are open, a global understanding of the molecular structure of Cs$_2$ is required. Potential curves of representative molecular states of Cs$_2$ in the energy range between the electronic ground state and the first dissociative-ionisation threshold are presented in Fig.\,\ref{fig:9}. The three panels focus on different ranges of the internuclear separation. Panel (a) shows the short range, with covalently bound molecular states; panel (b) shows the long-range region highlighting the three different important dissociation thresholds ${\rm Cs}~6{\rm s}_{1/2} + {\rm Cs}~6{\rm s}_{1/2}$, ${\rm Cs}^+ + {\rm Cs}^-$, and ${\rm Cs}~6{\rm s}_{1/2} + {\rm Cs}^+ + {\rm e}^-$; finally, panel (c) presents the electronic structure of the Cs atom. At short range (see Fig.\,\ref{fig:9}a), the Rydberg states of Cs$_2$ have potential curves that resemble those of the X$^+~{^2}{\rm \Sigma}^+_{\rm g}$ and the A$^+~{^2}{\rm \Sigma}^+_{\rm u}$ states of Cs$_2^+$. Interactions between Rydberg states having a repulsive A$^+$ ion core and those having a X$^+$ ion core and the resulting avoided crossings are responsible for the appearance of double minima in several curves. This aspect of the electronic structure of Cs$_2$ is analogous to the situation encountered in molecular hydrogen (see, \emph{e.g.}, Refs.\,\cite{kolos1963,sharp1971,quadrelli1990a,ross1994a,ross1994b}). Another common characteristic feature of the electronic structure of Cs$_2$ and H$_2$ is the existence of an attractive ion-pair potential curve at long range, represented as a thick black line in Fig.\,\ref{fig:9}b. In the discussion of long-range states of Cs$_2$, it is useful to consider both the internuclear separation $R$ and the distance ($r \approx a_0 n^2$) of the Rydberg electron from the centre of charge of the ion core. A classification of the long-range states of Cs$_2$ on the basis of the relative magnitude of $R$ and $r$ is presented in Fig.\,\ref{fig:10}.

Along the bisecting red line in this figure, $R$ is equal to $r$. For large values of $n$ (and thus of $r$ and $R$), this case corresponds closely to the long-range Rydberg molecules discussed in section 2 because the Fermi-contact term (see Eq.\,\ref{eq:Vpseudo}) responsible for the binding is proportional to the electron probability density at the position of the ground-state Cs atom. This situation is also realised in ion-pair states if one imagines that the Rydberg electron is ``attached'' to the ground-state atom in the 6s$_{1/2}$ state, forming the Cs$^-$ $^1$S$_0$ anion. Long-range Rydberg molecules and ion pairs are therefore related: In the former (see Fig.\,\ref{fig:3}), the attractive interaction between the slow Rydberg electron and the ground-state Cs atom correlates their motion around the Cs$^+$ ion core, and leads to binding as well as to a partial charge separation. This charge separation has, in previous work, been linked to the existence of a permanent electric dipole moment in a homonuclear diatomic molecule\,\cite{li2011,booth2015}, although, of course, an antisymmetrised nuclear wavefunction of the two Cs nuclei makes this dipole moment vanish, both in long-range Rydberg molecules and in Cs$^+$Cs$^-$ ion pairs.

At low values of $n$, the molecular states with $R \approx r$ are less isolated from other molecular states and the interactions cannot be conveniently treated with scattering theory because the kinetic energy of the electron markedly deviates from zero and more partial waves need to be included. Qualitative aspects are, however, preserved\,\cite{lipson1999,Kleimenov2007}.

The long-range Cs$_2$ molecules lie energetically just below the first ionisation energy of Cs ($E_{\rm I}({\rm Cs})/hc=31406.4673225(14)$\,cm$^{-1}$\,\cite{deiglmayr2016a}). Because the electron affinity, $E_{\rm EA}$(Cs), of Cs is 3804\,cm$^{-1}$\,\cite{scheer1998}, the ion-pair dissociation limit is located at the position of $n^* \approx 5.4$ Rydberg states. Long-range Rydberg molecules of the type discussed in section\,2 are therefore located in the ion-pair dissociation continuum into which they could, in principle, decay. The kinetic-energy release would roughly correspond to $E_{\rm EA}$(Cs) and thus to a de Broglie wavelength of $\sim0.1$\,$a_0$ for the ion-pair dissociation products. The overlap of the vibrational wavefunction of the long-range Rydberg molecule (see, \emph{e.g.}, Fig.\,\ref{fig:3}) with the wavefunction of the dissociated ion-pair is therefore expected to be extremely small. Consequently, the decay of the long-range Rydberg molecules into ion-pair states is likely to be of minor importance.

\begin{figure}[t!]
\center
\resizebox{0.60\linewidth}{!}{%
 \includegraphics{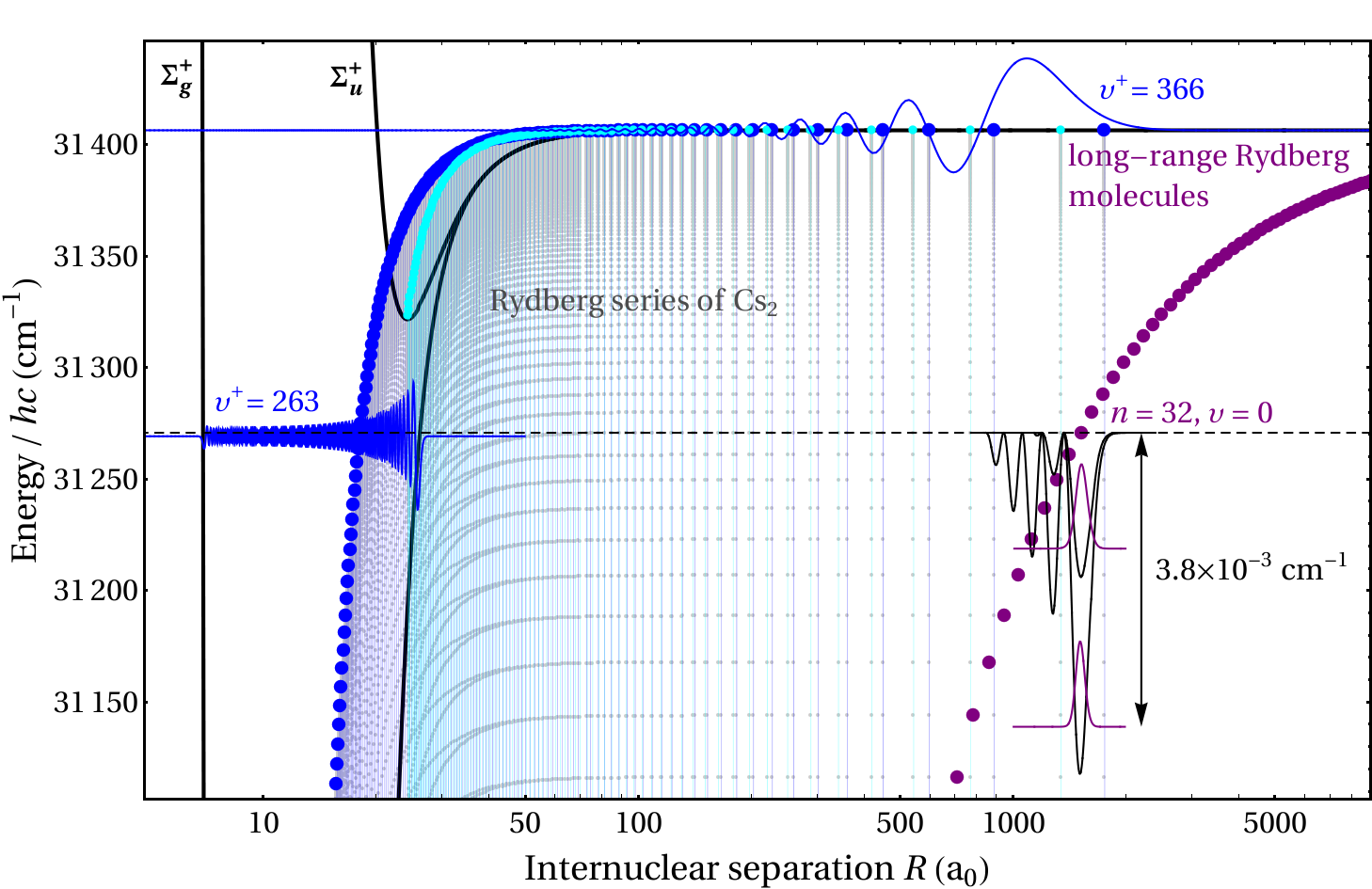} }
\caption{Comparison of energetic and structural characteristics of long-range Rydberg molecules and Cs$_2^+$ ions in high vibrational states. The molecular potentials of the X$^+ {^2}{\rm \Sigma}_{\rm g}^+$ and A$^+ {^2}{\rm \Sigma}_{\rm u}^+$ states of Cs$_2^+$ are given as solid black lines (taken from\,\cite{Jraij2005}). The energies and the expectation values $\left<1/R^2\right>^{-1/2}$ of the vibrational levels of the X$^+$ and the A$^+$ states are marked as dark and light blue dots, respectively. The Rydberg series that converge to each of these $v^+$ levels are schematically indicated as vertical thin dashed lines. The vibrational wavefunctions of the $v^+=263$ and $v^+=366$ levels of the X$^+$ state are drawn as thin dark blue lines. The energies and internuclear separations of the long-range Rydberg molecules discussed in section 2 are given as purple dots. The molecular potentials correlated to the Cs~$32$p$_{3/2} + {\rm Cs}~6$s$_{1/2}$ dissociation asymptote calculated with our Fermi s-wave scattering model are depicted on an energy scale magnified by a factor of 35000 and the two vibrational ground-state wavefunctions are presented as thin purple traces.}
\label{fig:11}
\end{figure}

The main decay product of the long-range Rydberg molecules detected in our experiments is the molecular Cs$_2^+$ ion. To illustrate possible mechanisms for the formation of Cs$_2^+$, Fig.\,\ref{fig:11} focuses on the energy region just below the first ionisation threshold  of Cs$_2$. The X$^+ {\rm \Sigma}_{\rm g}^+$ and A$^+ {\rm \Sigma}_{\rm u}^+$ potential curves of Cs$_2^+$, taken from\,\cite{Jraij2005}, are shown as solid black traces. Vibrational levels in these potentials were calculated using the mapped Fourier-grid technique\,\cite{Kokoouline1999} and their energies as a function of the expectation value ${\left<1/R^2\right>}^{-1/2}$ are indicated as blue and light blue dots for the ${\rm \Sigma}_{\rm g}^+$ and ${\rm \Sigma}_{\rm u}^+$ states, respectively. The wavefunctions of the $v^+=263$ and the $v^+=366$ levels of the ${\rm \Sigma}_{\rm g}^+$ state are shown as thin blue lines. The molecular potentials have their minima at relatively short ranges, but exhibit a long-range $\alpha/R^4$ scaling and support a few long-range states that have their peak probability amplitude at internuclear distances $>200a_0$. Molecular Rydberg series, schematically represented by vertical lines in Fig.\,\ref{fig:11}, converge to each of the ionic levels of Cs$_2^+$. The Rydberg states of Cs$_2$ have potential functions, and thus vibrational wavefunctions, that closely resemble those of Cs$_2^+$. The rotational structure adds significantly to the already very high molecular state density of Cs$_2$ and Cs$_2^+$ just below the first ionisation energy of Cs, but is omitted here for the sake of clarity.

The energies and internuclear separations of the Cs$_2$ long-range Rydberg molecules near Cs $n$p$_{3/2}$ + Cs 6s$_{1/2}$ dissociation asymptotes are presented as purple dots in Fig.\,\ref{fig:11}. The black lines represent the potentials calculated with our s-wave scattering model at $n=32$. These potentials are shown on an expanded energy scale for visibility and the vibrational wavefunctions of the two vibrational ground states are depicted as purple lines.
A multitude of Cs$_2^+(v^+) + {\rm e}^-$ channels are open at the position of the long-range Rydberg molecular states observed experimentally. However, the overlap of the vibrational wavefunctions of the long-range Rydberg molecules with those of accessible Cs$_2^+(v^+) + {\rm e}^-$ ionisation continua is also extremely small. It is conceivable that the decay of the long-range Rydberg molecules into Cs$_2^+$ occurs through sequential couplings to Rydberg states of Cs$_2$ in successive $\mathrm{\Delta} v=-1$ steps starting with the highest ($v^+=366$) level, which has a good Franck-Condon overlap with the long-range Rydberg molecular state. Channel interactions in molecular Rydberg states with $\mathrm{\Delta} v=-1$ are, indeed, known to be the strongest (see, \emph{e.g.}\,\cite{herzberg1972}). The decay of the long-range Rydberg molecules into Cs$_2^+ + {\rm e}^-$ may also arise from $l \geq 1$ contributions in the e$^-$-Cs scattering\,\cite{hamilton2002,bendkowsky2010}, which might help bridging the internuclear separation. We are currently extending our potential model of the long-range Rydberg molecules by including p-wave scattering contributions to the interaction Hamiltonian.

Rydberg-atom-pair states are electronically doubly excited and are located close to the dissociation threshold of two Cs$^+$ ions, \emph{i.e.}, far above the Cs(6s$_{1/2}$)+Cs$^+$ threshold. In our experiments, no molecular Cs$_2^+$ but only Cs$^+$ ions are observed in the decay of the Rydberg-atom-pair resonances. These resonances are embedded in a large number of Cs$^+ + {\rm Cs}^*$ dissociative-ionisation continua and the decay can be formally thought of as the decay of one of the interacting Rydberg atoms to a lower state and the ionisation of the other atom, the energy between the two atoms being exchanged \textit{via} dipole-dipole coupling\,\cite{li2005,viteau2008,tanner2008,vincent2014,pohl2003,robicheaux2014}. However, even though energetically allowed, the exact mechanisms leading to Cs$^+$ are not fully understood yet. The rates of direct dipole transitions between two Rydberg atoms at long range, which leaves one atom in a lower Rydberg state and the other one ionised, are too small to explain the experimental observations\,\cite{amthor2009}. The rates become significantly larger at distances smaller than the LeRoy radius\,\cite{leroy1973} where orbital overlap starts to play a role and Auger-type ionisation processes become possible\,\cite{kiffner2015}. If the Rydberg-atom pairs are excited to attractive potential curves, such short internuclear distances are reached typically within $\sim1$\,$\upmu$s because the two Rydberg atoms are attracted towards each other and eventually collide. We have modelled this process quantitatively in Cs Rydberg-atom pairs\,\cite{sassmannshausen2015}. Ionisation after such a collision of two highly excited atoms is known as Penning ionisation\,\cite{Penning1927} and has been identified as one important mechanism for spontaneous ion formation in ultracold Rydberg gases, which is the first step in the evolution of a dense ultracold Rydberg gas into a cold plasma\,\cite{robinson2000}.

\section{Conclusions}
We have observed long-range Rydberg molecules and Rydberg-atom-pair states in an ultracold gas of Cs atoms as sharp resonances close to atomic $n$p$_{3/2} \leftarrow 6$s$_{1/2}$ transitions. To describe the long-range Rydberg molecules, we have developed an s-wave scattering model accounting for the experimentally observed vibrational ground states in purely triplet scattering and singlet/triplet-mixed scattering potentials. Fitting the values of the singlet and the triplet zero-energy electron-Cs\,6s$_{1/2}$ s-wave scattering lengths $A_{\rm S,0}$ and $A_{\rm T,0}$ to reproduce the experimental binding energies has enabled the first determination of $A_{\rm S,0}$ based on experimental data. The observation of molecular resonances with even larger detunings than the vibrational ground states of the s-wave scattering potentials indicates the importance of p-wave scattering in Cs$_2$ long-range Rydberg molecules.

The spectral positions, line shapes and intensities of Rydberg-atom-pair resonances were compared to the results of calculations based on a long-range-interaction model describing molecular potentials associated with Cs$~nl + {\rm Cs}~n'l'$ dissociation asymptotes. The model reproduces the observed resonances with an accuracy of better than 5\,MHz without adjustable parameters. To achieve this accuracy, a large basis set of atom-pair states had to be considered and the importance of terms beyond the dipole-dipole contribution in the multipole expansion of the interaction was firmly established. Several resonances were found to originate from dipole-quadrupole interactions and terms up to the octupole-octupole term had to be taken into account to reproduce all experimental results. The same potential model has been successfully applied to describe the line shapes and broadenings in millimetre-wave transitions between interacting Rydberg-atom-pair states and the Penning ionisation of two interacting Rydberg atoms (see Ref.\,\cite{sassmannshausen2015}). Because the model adequately describes the behaviour of the interacting Rydberg gas at densities where two-body interactions dominate, it was also used to calculate the excitation probability as a function of the distance of two Rydberg atoms and the results were discussed in relation to an excitation-blockade radius. The limitations of simple blockade models were found to be particularly severe in the case of $n$p$_{3/2} \leftarrow 6$s$_{1/2}$ excitations.

Antiblockade effects and a very strong enhancement of the excitation probability were observed following two-colour two-photon excitation to Rydberg states, and the role played by ``seed'' Rydberg atoms was characterised in experiments monitoring both the Rydberg-excitation enhancement factors and the corresponding counting statistics.

\section*{Acknowledgements}
We gratefully acknowledge the support provided by the EU Initial Training Network COHERENCE under Grant No. FP7-PEOPLE-2010-ITN-265031 and by the Swiss National Science Foundation under Project No. 200020-14675 and the NCCR QSIT. We acknowledge the European Union H2020 FET Proactive project RySQ (grant N. 640378).

\end{document}